\documentclass[lettersize,journal]{IEEEtran}
\usepackage{amsmath,amsfonts}
\usepackage{array}
\usepackage[caption=false,font=normalsize,labelfont=sf,textfont=sf]{subfig}
\usepackage{textcomp}
\usepackage{stfloats}
\usepackage{url}
\usepackage{verbatim}
\usepackage{graphicx}
\usepackage{cite}

\usepackage[boxruled,noend]{algorithm2e}
\usepackage{amssymb}
\usepackage{amsthm}
\usepackage{float}  
\usepackage{svg} 
\usepackage{caption}
\usepackage{subcaption}
\usepackage{booktabs} 

\usepackage[hidelinks]{hyperref}
\usepackage{orcidlink}

\DeclareUnicodeCharacter{2061}{}
\DeclareUnicodeCharacter{2212}{}

\newtheorem{definition}{Definition}
\newtheorem{exmp}{Example}[section]
\newtheorem{theorem}{Theorem}
\newtheorem{remark}{Remark}

\newtheorem{property}{Property}
\newtheorem{corollary}{Corollary}

\allowdisplaybreaks[4]

\begin{document}

\title{An Information-theoretic Security Analysis of Honeyword}

\author{Pengcheng Su\textsuperscript{\orcidlink{0009-0003-9895-4995}}, Haibo Cheng\textsuperscript{\orcidlink{0000-0001-6677-463X}}, Wenting Li\textsuperscript{\orcidlink{0000-0002-2613-8257}}, Ping Wang\textsuperscript{\orcidlink{0000-0002-8854-2079}}

\thanks{Pengcheng Su is with the School of Computer Science, Peking University, Beijing, 100091, China. (e-mail: pcs@pku.edu.cn)
Haibo Cheng and Ping Wang are with the National Engineering Research Center of Software Engineering, Peking University, Beiiing,100091,China, and with Key Laboratory of High Confidence Software Technologies, Ministry of Education (Peking University), China. (e-mail: hbcheng@pku.edu.cn, pwang@pku.edu.cn) Wenting Li is with the Beijing Institute of Graphic Communication, Beijing, 102600, China. (e-mail: wentingli@pku.edu.cn) Haibo Cheng, Wenting Li and Ping Wang are the corresponding authors.
}
}

\markboth{}
{Su \MakeLowercase{\textit{et al.}}: An Information-theoretic Security Analysis of Honeyword and Password Probability Models}

\maketitle

\begin{abstract}
Honeyword is a representative ``honey'' technique that employs decoy objects to mislead adversaries and protect the real ones. 
To assess the security of a Honeyword system, two metrics—\textit{flatness} and \textit{success-number}—have been proposed and evaluated using various simulated attackers. 
Existing evaluations typically apply statistical learning methods to distinguish real passwords from decoys on real-world datasets. 
However, such evaluations may overestimate the system's security, as more effective distinguishing attacks could potentially exist.

In this paper, we aim to analyze the security of Honeyword systems under the \textit{strongest theoretical attack}, rather than relying on specific, expert-crafted attacks evaluated in prior experimental studies. 
We first derive mathematical expressions for the flatness and success-number under the strongest attack. 
We conduct analyses and computations for several typical scenarios, and determine the security of honeyword generation methods using a uniform distribution and the List model as examples. 

We further evaluate the security of existing honeyword generation methods based on password probability models (PPMs), which depends on the sample size used for training. We investigate, for the first time, the \textit{sample complexity} of several representative PPMs, introducing two novel polynomial-time approximation schemes for computing the total variation between PCFG models and between higher-order Markov models. These algorithms also hold independent theoretical significance. Our experimental results show that for small-scale password distributions, sample sizes on the order of millions—often tens of millions—are required to reduce the total variation below 0.1, even when flatness remains relatively high. 

A surprising result is that we establish an equivalence between flatness and total variation, thus bridging the theoretical study of Honeyword systems with classical information theory. Finally, we discuss the practical implications of our findings.

\end{abstract}


\section{Introduction}
\IEEEPARstart{P}{assword}-based authentication (PBA) remains widely used in computer systems. 
As a result, password storage on authentication servers continues to be a primary target for attackers and is frequently subject to data breaches. 
As of 21 April 2025, there have been 4,439 known data breaches, exposing more than 28.0 billion records \cite{44}. 
Since traditional PBA mechanisms lack the ability to detect password database breaches in a timely manner, such incidents are often discovered months or even years after they occur \cite{1}. 
This substantial time gap provides attackers with ample opportunity to recover plaintext passwords and compromise user accounts, posing serious security risks.

To detect password breaches, a novel and effective solution called the Honeyword system was proposed \cite{9}. 
In this scheme, the authentication server stores not only the user's real password but also $k-1$ plausible decoy passwords. 
These decoy passwords are referred to as \emph{honeywords}, and each of the $k$ passwords—one real and $k-1$ decoys—is collectively referred to as a \emph{sweetword}. 
If the password database is leaked, an attacker obtains $k$ sweetwords instead of a single real password\footnote{While sweetwords are typically hashed before being stored on the server, previous research assumes that the attacker can recover the plaintext of all sweetwords from their hash values \cite{1}, and this assumption is maintained in our work. We assume that the attacker has access to unlimited computational resources, and the challenge of the attack lies in the probabilistic nature of the Honeyword system.}, making it difficult to identify the correct one. 
Furthermore, if a honeyword is used during a login attempt, the system can raise an alert, since a legitimate user would not know which of the stored sweetwords are decoys. 

The leakage-detection capability of Honeyword systems relies on the indistinguishability between real passwords and honeywords. 
Ideally, with a perfect honeyword generation algorithm, no attacker should be able to identify the real password among the $k$ sweetwords with probability greater than $\frac{1}{k}$. 
However, in practice, accurately modeling the real password distribution is challenging, often resulting in statistical differences between real passwords and generated honeywords. 

Previous studies \cite{1,21,40} have employed statistical learning methods to construct sophisticated distinguishing attacks for evaluating the security of honeyword systems using real-world datasets. 
However, these attacks are experimental in nature and may overestimate the system's security, as more effective attacks could potentially exist.

In this paper, we investigate the security of honeyword systems from a theoretical perspective by analyzing the \textit{strongest possible attack}, which achieves the maximum distinguishing advantage allowed by the available information. 
We focus on two key security metrics:
\begin{enumerate}
\item \textit{Flatness}, which measures the average proportion $y$ of compromised accounts when the attacker is allowed $x$ guesses per account.
\item \textit{Success-number}, which quantifies the number of successfully compromised accounts $y$ after $x$ failed login attempts (i.e., honeyword activations) across a website containing $n$ accounts.
\end{enumerate}
These metrics have been used in previous studies with expert-designed attacks \cite{1,9}, but we evaluate them under the theoretically optimal attack strategy. 

We first analyze these two metrics from a theoretical perspective under the strongest attacker, rather than relying on specific, sophisticated attack strategies. 
By employing the classical likelihood ratio technique from information theory, we derive analytical expressions for both \textit{Flatness} and \textit{Success-number} under the optimal attack. 
This enables a rigorous information-theoretic analysis of honeyword system security. 
Notably, the mathematical structure of \textit{Flatness} is particularly well-defined, allowing exact computation in various representative scenarios.

Building upon the derived expression for \textit{Flatness}, we provide theoretical insights into the uniform honeyword generation method and offer explanations for empirical observations and conjectures reported in prior work. For instance, when the underlying password distribution follows a $0.7$-Zipf law, using uniform honeywords requires \( k = 175 \) to achieve a flatness level of 0.2.
We also establish the asymptotic formula $\epsilon_k(1) = \Theta\left(\frac{a}{k}\right) + b$, where $y=\epsilon_k(x)$ represents the Flatness curve and $a$ and $b$ are constants determined by the distributions of real passwords $P$ and honeywords $Q$. 
This result theoretically confirms a conjecture by Wang et al., originally inferred from experimental data \cite{21}. 
Furthermore, our analytical expression for \textit{Success-number} enables its exact evaluation, in contrast to prior studies that relied on Monte Carlo simulations to approximate its value. 

Furthermore, we evaluate the security of sample-based honeyword generation methods through a \textit{sample complexity} analysis. 
These methods rely on password probability models (PPMs)—such as the List Model, Markov Model, and PCFG Model—to learn the distribution of real passwords from real-world datasets and generate honeywords by sampling from the learned models. 
The security of such methods critically depends on their \textit{sample complexity}, which characterizes the number of samples (i.e., dataset size) required for the model to approximate the true password distribution.

In our sample complexity analysis, we compute the Flatness and \textit{total variation distance} (TV) for two PCFG models and two higher-order Markov models under varying sample sizes. 
The TV is a standard cryptographic and information-theoretical metric for measuring the divergence between two probability distributions. 
Calculating TV for password models is challenging due to its exponential computational complexity. 
To address this, we extend an existing TV estimation method \cite{33} and develop two fully polynomial-time approximation schemes (FPTAS) for such PPMs. 
Leveraging these algorithms, we conduct extensive experiments to study the sample complexity of PPMs in practical settings. Experimental results show that for these typical PPMs, achieving a total variation distance of 0.1 typically requires training data that is approximately ten times the size of the password distribution. 

In our analysis, we derive a mutual bound between the Flatness and TV, which may have independent theoretical value beyond the context of honeyword systems. 
For two probability distributions $P$ and $Q$, 
denote $Flat_k(P||Q) = \epsilon_k(1) - \frac{1}{k}$ and $\Delta_{TV}(P,Q)=\frac{1}{2}\sum_{x}|P(x)-Q(x)|$, the following relationship holds:
$$\frac{1}{k}\Delta_{TV}(P,Q) \le Flat_k(P||Q) \le \Delta_{TV}(P,Q).$$
This result links the study of Honeyword with classical information-theoretic theory. It indicates that in order to control the success rate of the strongest attacker's distinguishing attack, the TV between the password and honeyword distributions \textit{must} be sufficiently small.

Our main contributions are listed below: 
\begin{itemize}
\item \textbf{Mathematical Expressions of Honeyword Security Metrics:} We formalize the definitions of the two Honeyword security metrics and provide their precise mathematical expressions for the strongest attacker. Using the above theoretical results, we have conducted calculations for several representative scenarios.

\item \textbf{Equivalence of Flatness and Total Variation:}  The difference between flatness and $\frac{1}{k}$ can serve as an (asymmetric) distance metric between two probability distributions $P$ and $Q$. We prove that for any given $k$, this distance metric is equivalent to the total variation distance, i.e., $Flat_k(P||Q) = \Theta(\Delta_{TV}(P,Q))$. This theoretical result connects the Honeyword problem with classical information theory.

\item \textbf{Algorithms for Total Variation of PCFG Models and Markov Models:} We have developed polynomial-time approximation algorithms for calculating the total variation distance between two PCFG models and between two Markov models, which have both theoretical and practical significance.

\item \textbf{Sample Complexity of Password Probability Models:} Using the newly developed algorithms, we conduct experimental studies to evaluate the sample complexity of the List, PCFG, and Markov models. Our experimental results provide insights into the sample size required for PPM training in practice, as well as the fitting error and convergence rate.
\end{itemize}

\section{Background}\label{sec2}
In this section, we first provide detailed explanations on the detection strategy of Honeyword System. Then we review existing research results in the two major questions in Honeyword research: how to generate and attack honeywords \cite{21}.

\subsection{Detection Strategy}
A typical detection strategy, defined by two parameters $T_1$ and $T_2$, is described in \cite{1}. Specifically, an account will be locked if a certain number of honeywords (e.g., $\ge T_1$, with $T_1 = 3$) are submitted within a short time. Additionally, if a large number of honeywords (e.g., $\ge T_2$, with $T_2 = 10^4$) across the entire system are submitted within a short period, this could indicate a potential database breach. The analysis of these two thresholds corresponds to the two security metrics, flatness and success-number, respectively.

\subsection{Honeyword Generation Techniques}

According to a systematic review of Honeyword generation techniques conducted in 2022, more than 20 distinct techniques have been proposed \cite{10}. 
The mainstream honeyword generation techniques in current research are based on \textbf{password probability models}. These techniques generate honeywords by sampling from a learned password probability distribution \cite{23}. The concept of PPMs is a well-established area of research. A PPM $M$ aims to estimate the probability $P(pw)$ that a password $pw$ will be selected by a user. Furthermore, a trained password probability model $M$ can sample passwords according to its defined probability distribution $P_M(pw)$, which can then be used directly as honeywords.

We now introduce three representative password probability models: the \textit{List model}, the \textit{Markov model}, and the \textit{PCFG model} \cite{13}.

A \textbf{List model} directly utilizes the empirical distribution derived from the training set. Specifically, the probability of a password $P_M(pw)$ in the List model is determined by its frequency in the training set $S$: $$P_{List}(pw)=\frac{\#(pw)}{|S|},$$ where $\#(pw)$ denotes the number of occurrences of $pw$ in the set $S$, and $|S|$ represents the size of the training set $S$.
    
A \textbf{Markov model} treats the password generation process as a Markov process \cite{29}. Specifically, the probability of a password in a Markov model is
\begin{align*}
    P⁡_{Markov}(pw) &=\prod_{i=1}^{|pw|}\Pr ⁡[c_i\vert c_1 c_2\cdots c_{i-1}]\\
    &=\prod_{i=1}^{|pw|} \Pr⁡[c_i\vert c_{i-m}\cdots c_{i-1}],
\end{align*}
where $m$ is the order of the markov process \cite{29} and $|pw|$ is the length of $pw$. The conditional probability is set as its frequency in the training set. In other words, $$\Pr⁡[c_i\vert c_{i-m}\cdots c_{i-1}]=\frac{\#(c_{i-m}\cdots c_{i-1} c_i)}{\#(c_{i-m}\cdots c_{i-1})},$$ where $\#(w)$ is the frequency of occurrence of substring $w$ in the training set. 

In the field of password modeling, Markov models need to address the issue of varying password lengths. One approach, referred to as the ``Length Norm'' method, learns an empirical distribution of password lengths. Another approach introduces an ``End Symbol'' (`\textbackslash 0') to expand the character set, modeling the probabilities of various $n$-grams followed by the End Symbol based on the training dataset.
    
A \textbf{PCFG model} generates honeywords based on a probabilistic context-free grammar (PCFG) \cite{4}. The PCFG was originally introduced in natural language processing, and here we do not provide its general formal definition. The PCFG used in the password domain is of a ``two-layer'' structure, which is described as follows: The PCFG model parses the password into a sequence of ``tokens'' and assumes that the probability of a password is the product of the probability of its ``syntax'' and the probability of filling in the individual tokens.

Specifically, for a password like ``mice@123'', the probability is computed as:
$\Pr⁡[L_4|S_1|D_3]\cdot \Pr⁡[L_4\to``mice"]\cdot \Pr⁡[S_1\to``@"]\cdot \Pr⁡[D_3\to``123"]$, 
where $L_4$, $S_1$, and $D_3$ represent syntactic categories (letter, special character, digit). These probabilities are derived from the empirical frequencies observed in the training dataset:
\begin{itemize}
    \item $\Pr[L_4|S_1|D_3] = \frac{\#(L_4|S_1|D_3)}{|S|}$, where $\#(L_4|S_1|D_3)$ is the count of the observed syntax $L_4|S_1|D_3$ in the training set, and $|S|$ is the size of the training set $S$.

\item 
$\Pr[L_4 \to ``mice"] = \frac{\#(L_4 \to ``mice")}{\#(W_4)}$, where $\#(L_4 \to ``mice")$ is the number of times the token $L_4$ is filled with ``mice'', and $\#(L_4)$ is the total number of occurrences of the token $L_4$ in the training set.
\end{itemize}

\subsection{Existing Theoretic Results}\label{review2}
At CCS'13, Juels and Rivest introduced the Honeyword scheme and developed several techniques for generating honeywords \cite{9}. They also identified distinguishing attacks as potential threats to the Honeyword system. In their security analysis of distinguishing attacks, they introduced the concept of ``$\epsilon$-flat'' and used Bayesian analysis to determine the optimal attacking strategy, which was later formally articulated by Wang et al. at NDSS'18 \cite{1}.
Their main results are summarized as follows:

Assuming that the passwords of all users are independent and identically distributed (i.i.d.), the conditional probability that the $m$-th sweetword ($sw_{im}$) of the $i$-th user is its real password ($pw_i$) in a password file $F$ can be computed by considering only the sweetwords associated with the $i$-th user:
\[\Pr⁡[sw_{im}=pw_i \vert F]=\Pr⁡[sw_{im}=pw_i \vert sw_{i1},sw_{i2},\dots,sw_{ik}] .\]

Assuming that each honeyword of a user is also independent and identically distributed (iid), and letting $P$ denote the distribution of human passwords and $Q$ the distribution of honeywords, 
the conditional probability that $sw_{im}$ (the $m$-th sweetword of $i$-th User) is the real password of the $i$-th user can be calculated as follows:
\begin{align}
    \Pr⁡[sw_{im}=&pw_i|sw_{i1},\dots,sw_{ik}] \nonumber
    \\
    &=\frac{P(sw_{im})\prod_{j=1,j\neq m}^{k}Q(sw_{ij})}{\sum_{l=1}^{k}P(sw_{il})\prod_{j=1,j\neq l}^{k}Q(sw_{ij})} \nonumber
\\
&=\frac{\frac{P(sw_{im})}{Q(sw_{im})}}{\sum_{j=1}^{k}\frac{P(sw_{ij})}{Q(sw_{ij})}}. \label{f1}
\end{align}

This formula is crucial for subsequent analysis, and we provide a detailed explanation of its components. In the formula, $sw_{im}$ and $pw_i$ are two random variables, and the randomness originates from the sweetword list generation process (refer to the GenSWL procedure defined in Procedure \ref{GENSWL}). Given a user's $k$ sweetwords, the probability that the $m$-th sweetword $sw_{im}$ is the user's real password is proportional to its $P/Q$ value (i.e., $\frac{P(sw_{im})}{Q(sw_{im})}$). This is because the denominators in formula (\ref{f1}) are the same for all sweetwords of a user.

This result suggests that a user's password is most likely the sweetword with the highest $P/Q$ value, as this ratio reflects how well the sweetword fits the user’s actual password distribution $P$ relative to the honeyword distribution $Q$.

\section{Theoretical Calculation of Flatness}\label{sec_flat}
\subsection{Definition of Flatness Function}
To facilitate the next discussion, we'd like to present our formal definition of flatness.
\begin{definition}
    $FlatGame^{P,Q,k}_{\mathcal{A}}(n)$: Given a sweetword list consisting of one password (sampled from $P$) and $(k-1)$ honeywords (sampled from $Q$), the attacker $\mathcal{A}$ produces $n$ outputs (``guesses''). If the password is among the outputs, the attacker wins the game. Otherwise, the attacker fails. \label{def1}
\end{definition}

When $n=1$, the probability of success of the strongest attacker is $\max\limits_{\mathcal{A}}\Pr [FlatGame^{P,Q,k}_{\mathcal{A}}(1)=1]$, which is named \textit{flatness} (denoted by $\epsilon$) by Juels and Rivest \cite{9}. More generally, we define $\epsilon_k^{\mathcal{A}}(i)=\Pr[FlatGame^{P,Q,k}_{\mathcal{A}}(i)=1]$ and $\epsilon_k(i)=\max\limits_{\mathcal{A}}\epsilon_k^{\mathcal{A}}(i)$, where $1 \le i \le k$. We call $\epsilon_k(i)$ the \textbf{flatness function}. 

The pseudocode description of the above definition is provided in Game \ref{flat_game}, where \textit{GenSWL} (Generate Sweetword List) is a helper function. $\mathcal{PW}$ denotes the set of all possible passwords, i.e., the password space, and the notation $pw \gets_P \mathcal{PW}$ indicates that a password is sampled from the password space according to the probability distribution $P$.

\renewcommand{\algorithmcfname}{Procedure}

\begin{algorithm}[t]
  \SetAlgoLined

$pw \gets_{P} \mathcal{PW}$\\
$hw_1,hw_2,\cdots,hw_{k-1} \gets_{Q} \mathcal{PW}$\\
$SWL\gets [pw,hw_1,\cdots,hw_{k-1}]$\\
random\_shuffle$(SWL)$\\
\textbf{return} $(SWL,pw)$
  \caption{$GenSWL^{P,Q}(k)$}\label{GENSWL}

\end{algorithm}
\renewcommand{\algorithmcfname}{Game}
\begin{algorithm}[t]

$(SWL,pw) \gets GenSWL^{P,Q}(k)$\\
$Guesses=\{g_1,g_2,\cdots,g_n\} \gets \mathcal{A}(SWL)$\\
\textbf{return} $pw \in Guesses$.
  \caption{$FlatGame^{P,Q,k}_{\mathcal{A}}(n)$}\label{flat_game}

\end{algorithm}

In accordance with Kerckhoffs' principle, we assume the attacker has complete knowledge of the honeyword distribution $Q$. To calculate $\epsilon(i)$, we must consider the strongest attacker, who has full knowledge of both the real password distribution $P$ and the honeyword distribution $Q$. This assumption allows us to model the worst-case security scenario for the honeyword system, as the attacker can access all relevant distributions and thus optimize its strategy for distinguishing the real password from the honeyword.

\subsection{Continuous Case}

In the following, we distinguish between the continuous and discrete cases based on whether $P$ and $Q$ represent probability distributions of continuous or discrete random variables. Essentially, apart from some technical details, the primary difference between the calculations in these two cases is that the continuous case uses integration, while the discrete case uses summation.

The continuous case is generally easier to handle mathematically, so we will begin by discussing this case and then extend the analysis to the discrete case. Although the random variables in the password space for Honeyword are discrete, our mathematical approach for the continuous case can be adapted to Honey techniques that protect continuously distributed objects. This flexibility broadens the applicability of our security analysis framework.

In the continuous case, for convenience, we still use $\mathcal{PW}$ to denote the set of all possible values of the continuous random variable, and $pw$ to refer to the object sampled from this set. This notation helps maintain consistency with the discrete case, allowing for an easier transition between the two scenarios.

As we will demonstrate, the core of the theoretical computation of flatness lies in an interesting shift in perspective. The crucial idea is to transition from considering the two probability distributions of passwords to focusing on the two probability distributions of the $P/Q$ value ($\frac{P(pw)}{Q(pw)}$) of a password $pw$. This change of viewpoint simplifies the problem and allows us to perform the necessary calculations more effectively.

Consider two cumulative probability distributions defined as 
\begin{subequations}
    \begin{equation}
        F(x)=\Pr_{pw \gets_P \mathcal{PW}}[\frac{P(pw)}{Q(pw)}\le x], \label{def_f}
    \end{equation}
    and
    \begin{equation}
        G(x)=\Pr_{pw \gets_Q \mathcal{PW}}[\frac{P(pw)}{Q(pw)}\le x]. \label{def_g}
    \end{equation}
\end{subequations}
$P/Q$ is referred to as the likelihood ratio, which has broad applications in information theory. Here, we focus on the distribution of the likelihood ratio. Denote their corresponding probability density functions by $f(x)$ and $g(x)$. 

We further introduce the following two notations\footnote{Strictly speaking, the maximum value $M$ may not exist (i.e., $M=+\infty$) in the continuous case. However, this issue can be addressed by taking the limit. In the discrete case, on the other hand, the maximum value $M$ always exists and is finite. For simplicity, in the following, we assume that $M$ is a finite value.}: $$f(+\infty):= \Pr_{pw \gets_P \mathcal{PW}}[Q(pw)=0],\ \ M:=\max_{x: Q(x)\neq 0}\frac{P(x)}{Q(x)}.$$

Recall that the optimal attacker's strategy is to choose the sweetword with the largest $P/Q$ value for the first attack. The key observation is that the probability distribution of the $P/Q$ value of a password sampled from $P$ is exactly the function $f$ defined above. For example, if the $P/Q$ value of the user's real password is $1.4$, then the attacker will succeed in the first guess if and only if the $P/Q$ values of all $(k-1)$ honeywords are less than $1.4$. The probability of this event is given by $f(1.4)\cdot G^{k-1}(1.4)$. If $f(+\infty)=0$, meaning we consider all possible $P/Q$ values from $0$ to $M$, the total probability of the attacker's success can be calculated by integrating or summing over the range of possible values. This leads to the formula:
\begin{align}
    \epsilon_k(1)=\int_{0}^{M} f(x)G^{k-1}(x)\mathrm{d}x \label{epsilon11}
\end{align}
where $G$ is the cumulative distribution function of $g$ and $M$ is the maximum value of $\frac{P(pw)}{Q(pw)}, pw \in \mathcal{PW}$.

If $f(+\infty) >0$, then $\epsilon_k(1)=\int_{0}^{M} f(x)G^{k-1}(x)\mathrm{d}x +f(+\infty)$. We discuss cases of $f(+\infty)=0$ first and delay the cases of $f(+\infty) >0$ in Section \ref{sec33}. 

Similarly, $\epsilon_k(i)$ can be calculated. By the definition of $\epsilon_k(i)$ and the optimal attack strategy, the strongest attacker wins in a $FlatGame^{P,Q,k}_{\mathcal{A}}(i)$ if and only if it happens that the number of honeywords with $P/Q$ values greater than the $P/Q$ value of the real password is less than $(i-1)$. 

\begin{theorem}
    When $f(+\infty)=0$, the flatness function in the Definition \ref{def1} in the continuous case has the formula:
    \begin{equation}
    \epsilon_k(i)=\sum_{j=1}^{i}\int_{0}^{M}\tbinom{k-1}{j-1} f(x)G^{k-j}(x)\left(1-G(x)\right)^{j-1}\mathrm{d}x.
    \label{f6}
\end{equation}    
\end{theorem}

\begin{proof}
The strongest attacker guesses successfully in the $j$-th guess, which is equivalent to having exactly $(j-1)$ honeywords whose $P/Q$ value is greater than the $P/Q$ value $x$ of the real password. It is also equivalent to selecting $(j-1)$ honeywords from $(k-1)$ ones. Their $P/Q$ values are greater than $x$, and the remaining $(k-j)$ honeywords have $P/Q$ values less than $x$. 
This integral traverses all cases of $x$, taking into account all the possible configurations of honeywords that lead to a successful guess at the $j$-th attempt. The final result is obtained by summing $j$ from 1 to $i$.
\end{proof}

The following property plays an important part in the analysis of $\epsilon_k(i)$:

\begin{property}\label{prop0}
$f$ and $g$ defined above have the following relationship:
    \begin{align}
    f(x)=x\cdot g(x). \label{f2}
    \end{align}
\end{property}

Formula (\ref{f2}) has also been derived in theoretical works from other fields \cite{15}. To aid understanding, we present a calculation in the discrete case, which is primarily for conceptual clarity and can be easily extended to continuous scenarios.
Recall the definitions of the two probability distributions, $f$ and $g$. Consider a total of $m$ passwords $pw_1, pw_2, \dots, pw_m$, each having a $P/Q$ value denoted by $x$. Then, we can express $f(x)$ as follows: \[f(x)=\sum_{i=1}^m P(pw_i)=\sum_{i=1}^m x\cdot Q(pw_i)=x\sum_{i=1}^m Q(pw_i)=x\cdot g(x).\]
Table \ref{table1} shows the idea. In that case, $f(0.5)=0.05+0.01+\cdots$, in the meantime, $g(0.5)=0.1+0.02+\cdots$. It can be seen that $f(0.5)=0.5\times g(0.5)$.

\begin{table}[t]
\setlength{\abovecaptionskip}{0.5cm}
    \centering
    \caption{An example for $P/Q$ table in the password space}
    \begin{tabular}{ccccccc}
    \toprule
        $pw$ & 123456 & aaa123 & hello223  & p@ssword & ...\\
        \midrule
        $P(pw)$ & \textbf{0.05} & 0.015 & \textbf{0.01}  & 0.04 & ...\\

        $Q(pw)$ & \textbf{0.1} & 0.01 & \textbf{0.02} & 0.01 & ...\\

        $\frac{P(pw)}{Q(pw)}$ & \textbf{0.5} & 1.5 & \textbf{0.5} & 4.0 & ...\\
\bottomrule
    \end{tabular}
    \label{table1}
\end{table}

\begin{theorem}
When $f(+\infty)=0$, the $\epsilon(1)$ has the formula:
    \begin{equation}
        \epsilon_k(1)=\frac{1}{k}\left( M-\int_{0}^{M}G^k(x)\mathrm{d}x \right). \label{f3}
    \end{equation}
    
\end{theorem}

\begin{proof}
The key step is to use integrals by parts. We need to use property \ref{prop0} and the fact that $g$ is a derivative of $G$.
    \begin{align*}
    \epsilon_k(1)&=\int_{0}^{M} f(x)G^{k-1}(x)\mathrm{d}x\\
    &=\int_{0}^{M} x\cdot g(x)\cdot G^{k-1}(x)\mathrm{d}x\\
    &=\int_{0}^{M} x \cdot G^{k-1}(x)\mathrm{d}G(x)\\
    &=\frac{1}{k} \int_{0}^{M} x \cdot \mathrm{d}G^k(x)\\
    &=\frac{1}{k}\left(xG^k(x)|_0^M-\int_{0}^{M}G^k(x)\mathrm{d}x \right)\\
    &=\frac{1}{k}\left(M-\int_{0}^{M}G^k(x)\mathrm{d}x \right). \qedhere
\end{align*}
\end{proof}

Formula (\ref{f3}) demonstrates how $\epsilon_k(1)$ of a Honeyword system evolves with respect to $k$. Specifically, $\epsilon_k(1) = \Theta\left(\frac{M}{k}\right)$. The remaining term involves an integral of $G^k(x)$, where $G$ is the cumulative distribution function.

\begin{figure}[htbp]
    \centering
    \includegraphics[scale=0.4]{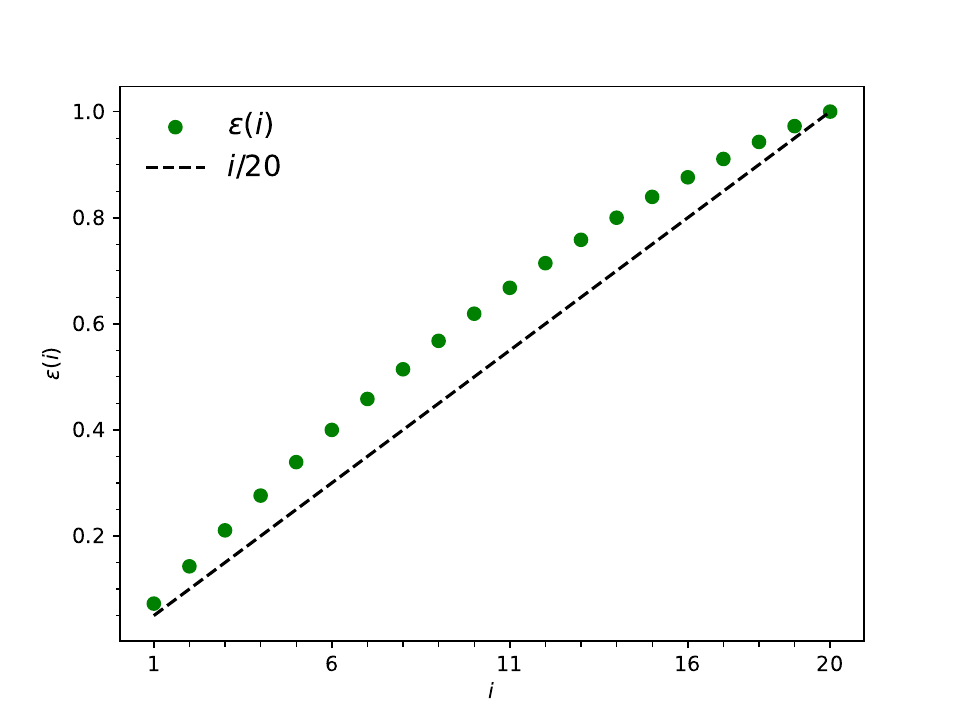}
    \caption{The flatness function in Example \ref{exmp1}}
    \label{fig3}
\end{figure}

\begin{exmp}\label{exmp1}
Consider these two continuous probability distributions defined on $[0,1]$:
\begin{align*}
    P(x)&=x+0.5, &0\le x \le 1\\
   Q(x)&=1, &0 \le x \le 1
\end{align*}
In this case, it can be calculated:
\begin{align*}
    f(x)&=x,\  &0.5\le x \le 1.5\\
    g(x)&=1,\  &0.5\le x \le 1.5 
\end{align*}
Therefore,
\begin{align*}
    \epsilon_k(1)&=\frac{1}{k}\left(M-\int_{0}^{M}G^k(x)\mathrm{d}x\right)=\frac{1}{k}(1.5-\frac{1}{k+1})\\
    &=\frac{1.5}{k}-\frac{1}{k(k+1)}.
\end{align*}
The $\epsilon_k(1)$ is $\Theta(\frac{1.5}{k})$.
We can also calculate all $\epsilon_k(i), 1 \le i\le k$:
\[\epsilon_k(i)=\frac{3k+2}{2k(k+1)}\cdot i -\frac{i^2}{2k(k+1)}.\]
It is a quadratic function of $i$.
Its function graph in the case of $k=20$ is shown in Figure \ref{fig3}.
\end{exmp}

\begin{remark}
The calculation given here is in the continuous case. This corresponds to the discrete case where the distribution of real passwords is a linear function and the distribution of false passwords is uniform. The continuous case can be regarded as the limit case where the size of the password space tends to infinity.
It is worth mentioning that $P(1)=1.5>1$ is because $P$ is the probability density function of a continuous random variable.
\end{remark}

\subsection{Discrete Case}
By interpreting integration as summation, we can extend the results from the continuous case to the discrete case. However, some technical aspects need to be addressed when generalizing to the discrete case. We directly provide the result for the discrete case, and the proof is given in Appendix \ref{prooftherm3}. (Due to space constraints, all appendices of this paper are included in the supplementary material.)

\begin{theorem}\label{therm3}
    The flatness defined in the Definition \ref{def1} in the discrete case has the formula:
    \[\epsilon_k(1)=\frac{1}{k}\sum_{x}x[G^{k}(x)-G^{k}(x^-)],\]
where $G(x^-)=\Pr_{pw \gets_Q \mathcal{PW}}[\frac{P(pw)}{Q(pw)}<x]$. That is, $G(x^-)$ is the left limit of the cumulative distribution function $G$ at x.
\end{theorem}

This is analogous to the result in the continuous case, which is given by the expression $\epsilon_k(1)=\frac{1}{k}\int_0^M x\mathrm{d}G^{k}(x)$.

Before moving on to the next example, we need to make some assumptions about the distribution $P$ of the real password. As we know, passwords generated by humans are not uniformly distributed. A series of studies have explored the distribution of human passwords. Empirical studies have shown that human passwords approximately follow a Zipf distribution \cite{11,12}. In the Zipf distribution model, the probability of the $r$-th largest password is given by
$\frac{r^{-\alpha}}{\sum_{i=1}^n i^{-\alpha}} $, where $\alpha$ is a parameter ($0<\alpha<1$) and $n$ is the size of the password space $\mathcal{PW}$ (i.e. $n=|\mathcal{PW}|$).
This implies that a small number of top-ranked passwords disproportionately contributes to the total probability mass.

\begin{figure}[h]
    \centering
    \includegraphics[scale=0.4]{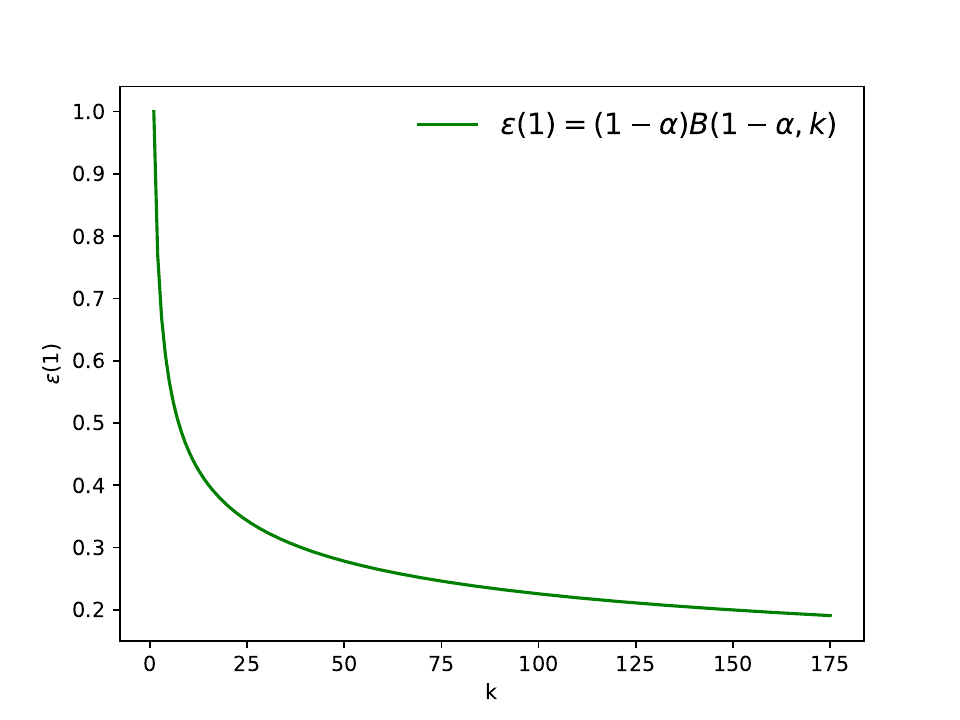}
    \caption{$\epsilon_k(1)$-$k$: Zipf password vs. uniform honeywords}
    \label{fig1}
\end{figure}

\begin{figure}[h]
    \centering
    \includegraphics[scale=0.4]{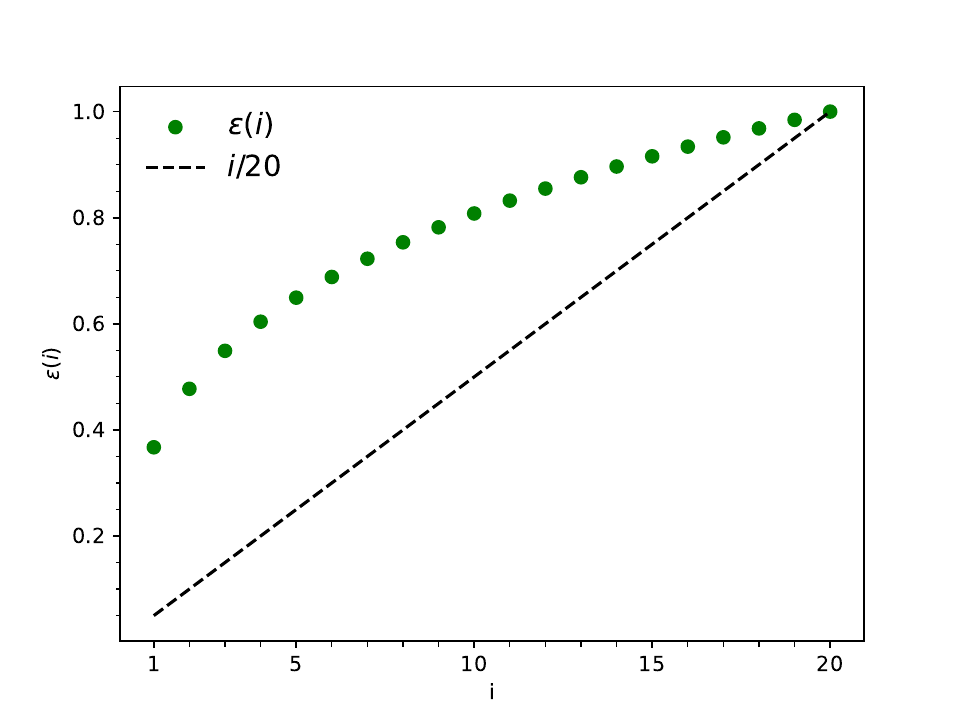}
    \caption{Flatness function ($k=20$): Zipf password vs. uniform honeywords}
    \label{fig111}
\end{figure}

\begin{exmp}\label{exmp2}
Consider that $P$ is a Zipf distribution with a parameter $\alpha$ over the password space, where the $i$-th largest probability is given by $P(pw_i)=\frac{i^{-a}}{\sum_{j=1}^n j^{-a}} $. $Q$ is a uniform distribution over the password space. Denote the sum of the Zipf probabilities by $S=\sum_{j=1}^n j^{-a} \approx \frac{1}{1-\alpha}n^{1-\alpha}$. The flatness can then be computed as follows:
\begin{align*}
    \epsilon_k(1)&=\frac{1}{k}\sum_{x}x[G^{k}(x)-G^{k}(x^-)]\\
    &=\frac{1}{k}\sum_{i=1}^{n}n\frac{(n-i+1)^{-\alpha}}{S}[(\frac{i}{n})^k -(\frac{i-1}{n})^k ]\\
    &\approx \frac{n^{1-\alpha}}{S}\sum_{i=1}^{n} \frac{1}{n} (1-\frac{i+1}{n})^{-\alpha}(\frac{i}{n})^{k-1}\\
    &\approx \frac{n^{1-\alpha}}{S} \int_0^1 (1-x)^{-\alpha}x^{k-1} \mathrm{d}x \\
    &=\frac{n^{1-\alpha}}{S} B(1-\alpha,k)\\
    &\approx (1-\alpha)B(1-\alpha,k),
\end{align*}
where $B$ is the beta function. In detail, $B(m,n)=\int_0^1 x^{m-1}(1-x)^{n-1} \mathrm{d}x$.
Supposed $\alpha=0.7$ \cite{21}, the $\epsilon_k(1)$-$k$ graph is shown in Fig.\ref{fig1}.

Similarly, we can calculate $\epsilon_k(i)=\sum_{j=1}^{i}(1-\alpha)\tbinom{k-1}{j-1}B(j-\alpha,k+1-j)$. The flatness function graph of $k=20$ is shown in Fig.\ref{fig111}

\end{exmp}

\subsection{Further Discussion \label{sec33}}
Consider a password $pw$ with $P(pw)>0$ and $Q(pw)=0$. When such a password occurs in a sweetword list, it is surely the real password because the honeyword generating technique will never generate it. Recall that $f(+\infty):=\Pr_{pw\gets_P \mathcal{PW}}[f(pw)>0\bigwedge Q(pw)=0]$.
In case of $f(+\infty)>0$,
\begin{align}
    \epsilon_k(1)&=\int_0^M f(x)G^{k-1}(x)\mathrm{d}x +f(+\infty) \nonumber\\
    &=\frac{1}{k}\left(M-\int_0^M G^{k}(x)\mathrm{d}x\right)+f(+\infty), \label{f4}
\end{align}
with a property:
\begin{align}
\int_0^M G(x)\mathrm{d}x &=xG(x)\vert_0^M -\int_0^M x\mathrm{d}G(x) \nonumber \\
&=M-\int_0^M f(x)\mathrm{d}x \nonumber \\
&=M-1+f(+\infty). \label{prop1}
\end{align}
This property is useful in the analysis in Section \ref{sec_crypto}.

It is clear from the formula (\ref{f4}) that $\epsilon_k(1) > f(+\infty)$. Some honeyword generation techniques have large $f(+\infty)$. The list model is a typical one.
The honeyword probability distribution $Q$ defined by the List model is the empirical distribution function derived from the training set $S$, which is sampled from the real password distribution $P$. In this model, any password that does not appear in 
$S$ has a probability of 0. Consequently, the probability mass for passwords outside the training set is zero. Therefore, $f(+\infty)=\sum_{pw\notin S}P(pw)$.

If $P$ is a uniform password distribution, a classical result is that when $\vert S \vert=\vert PW \vert =n$, $\mathbb{E}[f(+\infty)] \approx \frac{1}{e}$. The calculation is as follows:
\begin{align*}
    \mathbb{E}[f(+\infty)]&=\sum_{pw \in \mathcal{PW}}\frac{1}{n} \cdot \mathbb{E}[1_{pw  \notin S}]=\mathbb{E}[1_{pw  \notin S}]\\
    &=(1-\frac{1}{n})^n \approx \frac{1}{e}\approx 0.3679.
\end{align*}
More generally, an approximation formula for $f(+\infty)$ is given by $\mathbb{E}[f(+\infty)]\approx e^{-\frac{\vert S \vert}{n}}$. It shows that the expectation of $f(+\infty)$ is exponential decay with the size of the sample set. The attenuation constant is $\frac{1}{n}$.

How about the $f(+\infty)$ if $P$ is a Zipf password distribution with parameters $\alpha$ and $n$?
We can prove that the $f(+\infty)$ is also approximately exponential decay with the size of the sample set $|S|$.
The attenuation constant is $\Theta(\frac{1}{A})$, where $A=\sum_{j=1}^{n}j^{-\alpha}\approx \frac{n^{1-\alpha}}{1-\alpha}$. Notably, when $P$ is a Zipf distribution with $\alpha=0$, it reduces to the uniform distribution, and the results align with the earlier classical case. The detailed calculation process is included in Appendix \ref{appendix0}.

This result highlights an important consideration in the design of honeyword systems based on the List model. To ensure that $f(+\infty)$ of the List model small enough, the number of training samples must be on the order of $\Theta(n^{1-\alpha})$, where $n$ is the size of the password space and $\alpha$ is the parameter of the Zipf distribution. 
According to previous research, the Zipf parameters $\alpha$ for many human password datasets are between 0.6 and 0.9 \cite{11,12}.
This observation underscores the inherent difficulty of achieving a high level of security with the List model in honeyword generation. The experiments in Ref. \cite{1} also validate this theoretical analysis.

The above discussion concerns the sample complexity of the List Model with respect to the $f(+\infty)$ metric. In Section \ref{exp}, we will further explore the sample complexity of the List Model in relation to flatness and total variation.

\section{Theoretical Calculation of Success-number}\label{sec_sn}

In this section, we provide the formal definition of the success-number and derive its analytical expression. Due to the complexity of the mathematical techniques and calculations involved in the proof, we decide to include both a toy proof and the complete version in Appendix \ref{appendix_sn}. 

\subsection{Definition}
Another important security metric for the Honeyword system is the success-number. For the sake of simplicity in calculation, this paper focuses on the typical case where $T_1 = 1$, meaning the attacker is allowed only one attempt per account. We define the success-number based on a security game with a parameter $t$, which is designed to measure the number of vulnerable ``low-hanging fruits'' exposed to an attacker $\mathcal{A}$.

\begin{definition}
    $SNGame^{P,Q,k,U}_{\mathcal{A}}(t)$: Given sweetword lists of $U$ users (iid. generated by $GenSWL^{P,Q}(k)$ in Section \ref{sec_flat}) and the history of all guesses, the attacker $\mathcal{A}$ outputs a new guess $(id, pw^*)$ at each time, where $id \in \{1,2,\cdots, U \}$ and is not guessed before. Each guess is checked. If the $pw^*$ is the password of $id$-th user, the success number adds one. Otherwise, the failure number adds one. The game outputs the success number when the failure number reaches $t$, or after $U$ guesses are all checked. \label{def2}
\end{definition}

We are interested in the expected output of the $SNGame$. Let $\lambda_{U}^{\mathcal{A}}(t) = \mathbb{E}[SNGame^{P,Q,k,U}{\mathcal{A}}(t)]$ denote the expected success-number for a given attacker $\mathcal{A}$, and let $\lambda_{U}(t) = \max\limits_{\mathcal{A}} \lambda_{U}^{\mathcal{A}}(t)$ represent the maximum expected success-number across all possible attackers. In other words, $\lambda_U(t)$ is the expected success-number for the strongest attacker. We refer to $\lambda_U(t)$ as the \textbf{success-number function}.

\renewcommand{\thealgocf}{}

\renewcommand{\algorithmcfname}{Game}
\renewcommand{\thealgocf}{3}

\begin{algorithm}[t]

\For{$i=1$ to $U$}{$(L_i,index_i) \gets GenSWL^{P,Q}(k)$}

$F\gets [L_1,L_2,\cdots,L_U]$ \tcp{Files of $U$ accounts}

$SN\gets 0\ \ \ $ \tcp{Success-number}
$FN\gets 0\ \ \ $\tcp{Failure-number}
$V\gets \{1,2,\cdots,U\}\ \ \ $\tcp{Valid accounts}  
$history\gets [\ ]$  \tcp{History of $\mathcal{A}$'s attacks}
\While{$FN<t$ \textbf{and} $|V|>0$}
{$(id,guess) \gets \mathcal{A}(F,V,history)$\\ 
\If{ $id \notin V$}{\textbf{return} $-1$ \tcp{Invalid Attack}}
\ElseIf{$guess=index_{id}$}{$SN\gets SN+1$ \tcp{Successful attack}  $history.append( [id,guess,1] )$ }
\Else{$FN\gets FN+1$ \tcp{Failed attack} $history.append( [id,guess,0] )$}
$V\gets V- \{id \}$ \tcp{Account $id$ has been attacked}
}

\textbf{return} $SN$

\caption{$SNGame^{P,Q,k,U}_{\mathcal{A}}(t)$}\label{sn_game}
\end{algorithm}

\subsection{Theoretical Calculation}
Recall that we can use the Bayesian formula to calculate the maximum probability of a successful distinguishing attack on a sweetword list of $i$-th User [$sw_{i1},sw_{i2},…,sw_{ik}$]. On the condition that $T_1=1$, the attacker has only one opportunity for each account to submit a guess. Therefore, \[w_i=\max_{1\le m\le k}\frac{P(sw_{im})/Q(sw_{im})}{\sum_{j=1}^{k}P(sw_{ij})/Q(sw_{ij})} \] is exactly the probability of attacking this account successfully by the strongest attacker. The strongest attacker will calculate $w_i$ for each account using the formula (\ref{f1}) and then target the account with the largest $w$, the account with the second largest $w$, and so on.
 
Observe that $w_i$ is also a random variable. The randomness comes from $pw \gets_P \mathcal{PW}$ and $hw_1,\cdots,hw_{k-1} \gets_Q \mathcal{PW}$ in the generation of the sweetwords list. Denote the probability distribution of $w$ by $a(t)$, where $\frac{1}{k} \le t \le 1$. For example, $a(0.4)$ is the probability of generating a sweetword list in the face of which the optimal attacker has a success probability of 0.4.

\begin{theorem}\label{therm_lambda}
 $\lambda_U$-curve defined in Definition \ref{def2} has the formula:
\begin{equation*}
        \lambda_U(i)=U\sum_{j=1}^{i}\int_{\frac{1}{k}}^{1}t\cdot a(t)\tbinom{U-1}{j-1}(\psi(t) )^{U-j}(1-\psi(t))^{j-1}\mathrm{d}t ,
\end{equation*}
where
$$
\psi(t)=\int_0^t a(x)\mathrm{d}x +\int_t^1 x\cdot a(x) \mathrm{d}x .
$$
\label{therm4}
\end{theorem}

This result allows us to see the theoretical success-number graph for the first time, as opposed to some experimental curves resulting from numerous experiments. 
Like $\epsilon(1)$, the mathematical expression of success-number also reveals how $\lambda_U(i)$ evolves with $U$. The answer is that $U$ determines the convolution kernel of $\lambda_U(i)$.
\begin{theorem}\label{lambda_convolution}
Let $\phi(x)$ be the inverse function of $\psi(t)$ defined above. Then $\lambda_U$ has a formula:
    \begin{equation*}
        \lambda_U(i)= U\sum_{j=1}^{i}\int_{\frac{1}{k}}^{1} \frac{\phi(x)}{1-\phi(x)} \tbinom{U-1}{j-1} x^{U-j}(1-x)^{j-1}\mathrm{d}x.
    \end{equation*}
This formula can be viewed as a series of beta convolution kernels $\tbinom{U-1}{j-1} x^{U-j}(1-x)^{j-1}$ applied to a function $\frac{\phi(x)}{1-\phi(x)}$.
    
\end{theorem}

The proof is included in Appendix \ref{appendix_convolution}.

\section{The equivalence between Flatness and Total Variance}\label{sec_crypto}

Many cryptographic definitions involve scenarios similar to the following: the challenger samples an object $m_0$ from a probability distribution $P$ and an object $m_1$ from a probability distribution $Q$, then generates a random bit $b$ and sends the message $m_b$ to the attacker. The attacker is tasked with outputting a bit $b'$ to identify the distribution from which the message originated. Typically, we are concerned with the winning probability $\Pr[b' = b]$. It is well-known that the maximum winning probability achievable by any attacker is exactly $\frac{1}{2} + \Delta_{TV}(P, Q)$, where $\Delta_{TV}(P, Q)$ denotes the total variation distance between the distributions $P$ and $Q$.

What happens if the challenger sends both messages to the attacker and asks the attacker to determine which message comes from which distribution? Note that this precisely aligns with the definition of the flatness game in the Honeyword system with parameter $k=2$. In this case, the maximum winning probability for any attacker is essentially determined by the flatness $\epsilon_2(1)$.

From Flatness, we can derive a metric $Flat_k(P|| Q)$ for the distance between two probability distributions $P$ and $Q$, defined as $Flat_k(P|| Q) = \epsilon_k(1) - \frac{1}{k}$. When $P = Q$, $Flat_k(P|| Q) = 0$. It is an asymmetric distance metric: $Flat_k(P || Q) \neq Flat_k(Q || P)$.
The following theorem establishes the equivalence between these two distance metrics for probability distributions.
\begin{theorem}\label{crypto-game}
    $\frac{1}{k}\Delta_{TV}(P,Q) \le Flat_k(P||Q) \le \Delta_{TV}(P,Q)$
\end{theorem}
\begin{proof}
The proof of the first part of this theorem involves constructing a distinguishing attacker $\mathcal{A}$ such that its $\epsilon_k^{\mathcal{A}}(1)$ equals $\frac{1}{k} + \frac{1}{k} \Delta_{TV}(P, Q)$. Since $\epsilon_k(1)$ is defined as the upper bound on the success probability of all attackers, it follows that $\frac{1}{k} + \frac{1}{k} \Delta_{TV}(P, Q) \le \epsilon_k(1) = \frac{1}{k} + Flat_k(P || Q)$. The proof of the second part of the theorem relies on the analytical expression for flatness derived earlier, and several mathematical techniques are used to obtain the result. We present the full proof in Appendix \ref{appendix2}. \qedhere
\end{proof}

We have the following proposition:
\begin{corollary}
    $Flat_2(P,Q) \le \Delta_{TV}(P,Q) \le 2Flat_2(P,Q)$
\end{corollary}

This result establishes a surprising connection between the concept of flatness in Honeyword research and the total variation distance in classical information theory. Furthermore, it provides a simple and practical Monte Carlo algorithm for approximately estimating the total variation distance between any two probability models (e.g., neural networks). Since flatness comes with a natural Monte Carlo estimation algorithm, we can estimate the total variation distance between any two probability distributions $P$ and $Q$ by estimating $Flat_2(P, Q)$. 
This approach only requires that the two distributions $P$ and $Q$ are efficiently sampleable and that the corresponding sample probabilities can be computed. When the total variation distance between the two distributions is small, this estimate effectively bounds the total variation distance within a relatively narrow range.

\renewcommand{\algorithmcfname}{Algorithm}
\renewcommand{\thealgocf}{4}

\begin{algorithm}[t]
\KwIn{Two probability distribution $P, Q$, a parameter $k$, 
Monte Carlo experiment iterations $N$}
\KwOut{Estimated value of $Flat_k(P,Q)$ }
Construct the strongest attacker $\mathcal{A}^*$\\
\tcp{ Described in Section \ref{review2}}
$res\gets 0$\\
\For{$i=1$ to $N$}{
$res\gets res+FlatGame^{P,Q,k}_{\mathcal{A}^*}(1)$\\
\tcp{Repeat simulation of Game \ref{flat_game} }
}
\textbf{Return} $\frac{res}{N}-\frac{1}{k}$;
\caption{Monte-Carlo Algorithm for $Flat_k(P,Q)$}\label{algo_flatness}
\end{algorithm}

\section{Algorithm for Total Variation of PCFG Models and Markov Models}
Total variation distance is a classical metric used to measure the distance between two probability distributions, and it is widely applied in information theory and cryptography. Considerable theoretical work has been done on both exact and approximate algorithms for calculating the total variation distance between certain types of distributions \cite{34,35}. However, for many distributions, the exact computation or even estimation of the total variation distance between them has proven to be a difficult problem \cite{32,36}.

In the case of PPMs, the total variation distance between two list models is simply the total variation distance between their empirical distributions over the training set. This can be computed by enumerating each password in the training set to obtain the exact value. However, the probability distributions defined by Markov models and PCFG models are generalized from the training set to the entire password space according to their respective rules. Computing the exact total variation distance between these models would require enumerating the entire password space. Since the size of the password space grows exponentially with password length, such enumeration becomes computationally infeasible.

A recent breakthrough in total variation distance computation is the work by Feng et al. at SODA 2024. It provided fully polynomial-time approximation schemes (FPTAS) for calculating the distance between two product distributions and between two first-order Markov models \cite{33}. Building on their work, we develop FPTAS algorithms for calculating the total variation distance between two PCFG models and two higher-order Markov models. 

In Section \ref{feng etal}, we present the main results of Feng et al.'s work \cite{33}. Following this, we introduce a reduction from the problem of computing the total variation distance for PCFG models to the problem of computing the TV distance for product distributions. 
The algorithm for higher-order Markov models is relatively straightforward so
we briefly discuss some noteworthy aspects of our algorithm.

\subsection{Existing Work} \label{feng etal}
A product distribution is a probability distribution over a product space, where the random variables in the space are independent. Formally, let $X = (X_1, X_2, \dots, X_n)$ be a random vector, and let $X_i$ be independent random variables taking values in a finite set $[q]=\{1,2,\cdots,q\}$. A product distribution $\mathbb{P}$ over $[q]^n$ is defined by the product of marginal distributions of each $X_i$, such that:
\[
\mathbb{P}(X = (x_1, x_2, \dots, x_n)) = \prod_{i=1}^n \mathbb{P}_i(x_i),
\]
where $\mathbb{P}_i(x_i)$ is the marginal probability of $X_i$ taking the value $x_i$.

Feng et al.'s work presents the following results and corresponding algorithms \cite{33}.
\begin{theorem}[TV distance between product distributions]
Feng et al. proposed a deterministic algorithm such that given two product distributions $\mathbb{P}, \mathbb{Q}$ over $[q]^n$ and $\epsilon > 0$, it outputs $\widehat{\Delta}$ satisfying
\[
(1 - \epsilon) \Delta_{\text{TV}}(\mathbb{P}, \mathbb{Q}) \leq \widehat{\Delta} \leq \Delta_{\text{TV}}(\mathbb{P}, \mathbb{Q})
\]
in time 
\[
O\left(\frac{qn^2}{\epsilon} \log q \log \frac{n}{\epsilon \Delta_{\text{TV}}(\mathbb{P}, \mathbb{Q})}\right).
\]
\end{theorem}

In Appendix \ref{appedix_tv1}, we provide a very concise introduction to the steps and principles of their algorithm for the total variation distance of product distributions.

\begin{theorem}[TV distance between $n$-step Markov chains]
Feng et al. proposed a deterministic algorithm such that given two $n$-step Markov chains $\mathbb{P}, \mathbb{Q}$ in state space $[q]$ and $\epsilon > 0$, it outputs $\widehat{\Delta}$ satisfying
\[
(1 - \epsilon) \Delta_{\text{TV}}(\mathbb{P}, \mathbb{Q}) \leq \widehat{\Delta} \leq \Delta_{\text{TV}}(\mathbb{P}, \mathbb{Q})
\]
in time 
\[
O\left(\frac{q^2 n^2}{\epsilon} \log q \log \frac{n}{\epsilon \Delta_{\text{TV}}(\mathbb{P}, \mathbb{Q})}\right).
\]
\end{theorem}

The term $n$-step refers to the probability distributions defined over $[q]^n$, where $[q]$ denotes the state space of the Markov chains. A detailed explanation of this algorithm requires significant exposition. Interested readers are encouraged to consult their work for further details.

\subsection{Algorithm for Total Variation of PCFG Models}
The PCFG model is highly analogous to a product distribution.
Let $\mathcal{PW}|_S$ denote the set of passwords constrained by syntax $S=C_1|C_2|\cdots|C_m$.
It can be observed that given a fixed syntax $S$, the conditional probabilities $\Pr_M[pw|pw \in \mathcal{PW}|_S ]$ of a password PCFG model $M$ form a product distribution. It can be understood this way: the entire password space is partitioned into subspaces based on syntaxes, and a product distribution is defined over each subspace. The probability distribution defined by the PCFG model is a weighted combination of these product distributions, where the weighting coefficients are determined by the probability distribution over syntaxes $\Pr_{M}[S]$.

The aforementioned structure of the PCFG provides a significant advantage for calculating its total variation distance. For two PCFGs, their probability distributions overlap only on syntaxes that are shared between the two models. If a syntax $S$ is exclusive to one PCFG $M_i$, the corresponding password subspace contributes $\frac{1}{2}\Pr_{M_i}[S]$ to the total variation distance. If a syntax $S$ is shared by both PCFGs, we need to compute the sum:
\begin{align*}
&\frac{1}{2}\sum_{pw \in \mathcal{PW}|_S}\Big| \Pr_{M_1}[pw]-\Pr_{M_2}[pw]\Big|\\
&\ \ \ \ =\frac{1}{2}\sum_{pw \in \mathcal{PW}|_S}\Big| \Pr_{M_1}[S]\cdot \Pr_{M_1}[S \to pw]-\Pr_{M_2}[S]\Pr_{M_2}[S \to pw]\Big|
\end{align*}

In the above expression, $\Pr\limits_{M_i}[S \to pw]$ represents a product distribution (denoted by $M_i(S)=M_1(C_1)\times M_1(C_2) \times \cdots \times M_1(C_m)$), but $\Pr_{M_i}[S]$ is not. Fortunately, we can associate it with a product distribution by constructing two auxiliary probability distributions. Let us first consider the following intuitive example. 
Define two distributions $\mathbb{P}_0,\mathbb{Q}_0$ over $\{1,2,3\}$:
$$\mathbb{P}_0(1)=p,\ \ \ \ \mathbb{P}_0(2)=1-\mathbb{P}_0(1),$$
$$\mathbb{Q}_0(1)=q,\ \ \ \ \mathbb{Q}_0(3)=1-\mathbb{Q}_0(1).$$
Then, we consider the total variation distance between $\mathbb{P}_0 \times M_1(S)$ and $\mathbb{Q}_0 \times M_2(S)$:
\begin{align*}
&\Delta_{TV}(\mathbb{P}_0 \times M_1(S),\mathbb{Q}_0 \times M_2(S))\\
&=\frac{1}{2}\sum_{i=1}^3 \sum_{pw \in \mathcal{PW}|S}\Big|\mathbb{P}_0(i)\cdot M_1(S)[pw]-\mathbb{Q}_0(i)\cdot M_2(S)[pw]\Big|\\
&=\frac{1}{2}\sum_{pw \in \mathcal{PW}|_S}\Big| \Pr_{M_1}[S]\cdot M_1(S)[pw]-\Pr_{M_2}[S]M_2(S)[pw]\Big|\\
&\ \ \ \ \ \ \ +\frac{1}{2}[\mathbb{P}_0(2)+\mathbb{Q}_0(3)]\\
&=\frac{1}{2}\sum_{pw \in \mathcal{PW}|_S}\Big| \Pr_{M_1}[pw]-\Pr_{M_2}[pw]\Big|+\frac{1}{2}[\mathbb{P}_0(2)+\mathbb{Q}_0(3)]
\end{align*}

Thus, we can compute the total variation distance between two PCFGs over the subspace $\mathcal{PW}|_S$ by calculating $\Delta_{TV}(\mathbb{P}_0 \times M_1(S),\mathbb{Q}_0 \times M_2(S))$. The total variation distance between two product distributions can be computed using existing approximation algorithms, which guarantee a relative error of no more than $\epsilon$. By iterating through all shared syntaxes of the two PCFGs, summing the estimated results for each syntax, and adding the probabilities of syntaxes unique to each PCFG, we can obtain an estimated value for the TV distance between the two PCFGs. The pseudocode for this algorithm is in \ref{algo_pcfg}. It uses a more sophisticated selection of $\mathbb{P}_0$ and $\mathbb{Q}_0$ that can further reduce the final estimation error. We present the following theorem to formalize this improvement.

\renewcommand{\algorithmcfname}{Algorithm}
\renewcommand{\thealgocf}{5}

\begin{algorithm}[t]
\caption{Algorithm for Total Variation of PCFG Models}\label{algo_pcfg}
\KwIn{Two PCFG Models $M_1, M_2$, an error bound $\epsilon>0$}
\KwOut{Approximate TV $\hat{\Delta}(M_1,M_2)$ }
\textbf{External} $TV4ProductDistritbuion$ in \cite{33}\\

$res\gets 0$\\
\For{Each ``syntax'' $S=C_1|C_2|\cdots|C_m$ of $M_1$ and $M_2$}{
$p\gets \Pr_{M_1}[S]$\ \tcp{Prob of syntax $S$ in $M_1$}
$q\gets \Pr_{M_2}[S]$\ \tcp{Prob of syntax $S$ in $M_2$}

\If{$p=0||q=0$}{$res+=\frac{p+q}{2}$\ \ \tcp{Straightforward Computation} \textbf{continue} \\}

Define a distribution $\mathbb{P}_0$:\\
$\mathbb{P}_0(1):=\frac{p}{\max(p,q)},\ \mathbb{P}_0(2):=1-\mathbb{P}_0(1),\ \mathbb{P}_0(3):=0$\\
Define a distribution $\mathbb{Q}_0$:\\
$\mathbb{Q}_0(1):=\frac{q}{\max(p,q)},\ \mathbb{Q}_0(2):=0,\mathbb{Q}_0(3):=1-\mathbb{Q}_0(1)$\\
\For{$i = 1$ to $m$}{
Define two distributions:\\
$\mathbb{P}_i := M_1(C_i)$\tcp{Prob Distribution on $C_i$ in $M_1$}
$\mathbb{Q}_i := M_2(C_i)$\tcp{Prob Distribution on $C_i$ in $M_2$}
}
Define a product distribution $\mathbb{P}:= \mathbb{P}_0\cdot \mathbb{P}_1\cdot \mathbb{P}_2\cdots \mathbb{P}_m$\\
Define a product distribution $\mathbb{Q}:= \mathbb{Q}_0\cdot \mathbb{Q}_1\cdot \mathbb{Q}_2\cdots \mathbb{Q}_m$\\
$t\gets TV4ProductDistritbuion(\mathbb{P},\mathbb{Q},\epsilon/2)$\\
$res+= \max(p,q)\left( t-(1-\frac{p+q}{2}) \right)$\\
}
\textbf{return} $res$;
\end{algorithm}

\begin{theorem}\label{therm_pcfg}
The estimated output $\hat{\Delta}$ by Algorithm \ref{algo_pcfg} satisfies the following properties:
$$(1-\epsilon)\Delta_{TV}(M_1,M_2) \le \hat{\Delta} \le \Delta_{TV}(M_1,M_2).$$
\end{theorem}

The proof of the theorem is provided in Appendix \ref{appendix_tv2}.

\subsection{Algorithm for Higher-Order Markov Model}
Feng et al. designed an algorithm to calculate the TV distance between two first-order Markov chains \cite{33}. 
By treating $n$-grams as a new set of states, a $n$-order Markov process can be reduced to a first-order Markov process. However, the algorithm proposed by Feng et al. is only applicable to data of uniform length, whereas passwords exhibit varying lengths, necessitating additional techniques to handle this issue. 

The Markov Model with the End Symbol can directly address this issue at the cost of increasing the character set size by one. In contrast, when using the Length Norm Markov Model, calculating the total variation distance requires employing the aforementioned algorithm for PCFGs proposed in this paper. This is because the Length Norm Markov Model also exhibits a similar two-layer structure: it first computes the probability of a password length (analogous to the ``syntax'' in PCFG) and then multiplies it by the probability given by the corresponding Markov process (analogous to the multiplicative distribution in PCFG).

Furthermore, treating $n$-grams as a new character set results in a size of $q^n$ (where $n$ is the order of the Markov model). Since the time complexity of Feng et al.'s algorithm is proportional to the square of the character set size, the runtime of the algorithm increases exponentially as $n$ grows. Fortunately, the parameters of the password Markov model are derived from the training set, and its transition probabilities are sparse (many of them are zero).

\section{Experiment}\label{exp}
In this section, we conduct experiments on the sample complexity of the password probability model. We focus on the total variation and flatness between the target distribution and the probability distribution learned by the model.

\subsection{Ethical Statement}
This paper uses public password datasets commonly used in the password security research community. The purpose of this study is to deeply understand the security of the current password probability model and Honeyword mechanism. Our experimental research will not cause security issues.

\subsection{Settings}
Since the distribution of real human passwords does not necessarily follow the Markov Model or PCFG Model, even when the number of samples is infinite, we cannot guarantee that there is a Markov Model or PCFG Model that converges to this distribution. For the sample complexity problem of the password probability model, we use the following idea: Assuming that the target distribution is indeed a Markov/PCFG model, how many samples can make the trained Markov/PCFG model converge to it.

In particular, we use a password dataset $D$ to train a model $M$ as the target distribution of passwords. Sample a sample set $S$ from the target distribution, and train an estimated model $M'$ from $S$. We test the relationship between $\Delta_{TV}(M,M')$ and $Flat_k(M,M')$ and $|S|$.

Due to the randomness inherent in the sample set $S$ drawn from the target distribution in each sampling process, the distribution of the trained model $M'$ may vary. However, from the experimental results, the absolute error of the total variation distance and flatness among different $M'$ is around $10^{-3}$, which indicates that the sample size $|S|$ largely determines the convergence of the model, with the impact of sample randomness being minimal.

We conducted experiments using the 12306 dataset, the Yahoo dataset and the Duowan dataset (containing approximately $1.3 \times 10^5$, $4.4 \times 10^5$, $49.8 \times 10^5$ passwords, respectively). For the estimation of the total variation distance for PCFG and Markov models, we set the parameter in the estimation algorithm to 0.1, meaning that the relative error between the estimated value and the true value is less than 0.1. The Markov model we use includes an End Symbol and has an order of 3. To accelerate the estimation algorithm, we limit our experiments on the Markov model to passwords in the dataset with a maximum length of 15. Due to space limitations, the curves of the experimental results on the 12306 dataset are presented in the appendix \ref{appendix_experiment}.

\subsection{Experimental Results}
\begin{table*}[htbp]
    \centering
    \caption{Sample Complexity of PPMs (12306, $|D|=1.3\times 10^5$)}
    \begin{minipage}{0.32\textwidth}
        \centering
        \resizebox{\textwidth}{!}{
        \begin{tabular}{ccccccccc}    
    \toprule
       $|S|$/$10^5$ & 1 &3  &5  &10  & 20 & 30 & 50 & 100     \\ 
       \midrule
       $TV$  & 0.430 & 0.247 & 0.188 & 0.134 & 0.095 & 0.077 & 0.060 & 0.042    \\ 
       $Flat_2$  & 0.279 & 0.171 & 0.135 & 0.095 & 0.068 & 0.056 &0.043 & 0.030   \\ 
       $Flat_{20}$ & 0.425 & 0.146 & 0.091 & 0.052 & 0.032 & 0.024 &0.017 & 0.012   \\ 
    \bottomrule
    \end{tabular}
        }
        \caption*{List}
    \end{minipage}
    \hfill
    \begin{minipage}{0.32\textwidth}
        \centering
        \resizebox{\textwidth}{!}{
        \begin{tabular}{ccccccccc}    
    \toprule
       $|S|$/$10^5$ & 1 &3  &5  &10  & 20 & 30 & 50 & 100     \\ 
    \midrule
       $TV$  & 0.379 & 0.216 & 0.166 & 0.118 & 0.083 & 0.068 & 0.052 & 0.037    \\ 
       $Flat_2$  & 0.265 & 0.157 & 0.122 & 0.086 & 0.062 & 0.051 &0.038 & 0.027   \\ 
       $Flat_{20}$ & 0.360 & 0.128 & 0.082 & 0.048 & 0.029 & 0.023 &0.017 & 0.011   \\ 
    \bottomrule
    \end{tabular}
    }
        \caption*{PCFG}
    \end{minipage}
    \hfill
    \begin{minipage}{0.32\textwidth}
        \centering
        \resizebox{\textwidth}{!}{
        \begin{tabular}{ccccccccc}    
    \toprule
       $|S|$/$10^5$ & 1 &3  &5  &10  & 20 & 30 & 50 & 100     \\ 
    \midrule
       $TV$  & 0.385 & 0.219 & 0.176 & 0.127 & 0.097 & 0.084 & 0.072 & 0.061    \\
       $Flat_2$  & 0.260 & 0.157 & 0.124 & 0.092 & 0.071 & 0.061 &0.052 & 0.045   \\ 
       $Flat_{20}$ & 0.292 & 0.128 & 0.086 & 0.055 & 0.037 & 0.030 &0.025 & 0.020   \\ 
    \bottomrule
    \end{tabular}
    }
        \caption*{Markov}
    \end{minipage}
\end{table*}

\begin{table*}[htbp]
    \centering
    \caption{Sample Complexity of PPMs (Yahoo, $|D|=4.4\times 10^5$)}
    \begin{minipage}{0.32\textwidth}
        \centering
        \resizebox{\textwidth}{!}{
        \begin{tabular}{ccccccccc}    
    \toprule
       $|S|$/$10^5$ & 1 &3  &5  &10  & 20 & 30 & 50 & 100     \\ 
       \midrule
       $TV$  & 0.674 & 0.420 & 0.306 & 0.222 & 0.157 & 0.127 & 0.099 & 0.070    \\ 
       $Flat_2$  & 0.392 & 0.278 & 0.222 & 0.160 & 0.114 & 0.094 &0.073 & 0.051   \\ 
       $Flat_{20}$ & 0.668 & 0.408 & 0.266 & 0.131 & 0.072 & 0.052 &0.036 & 0.023   \\
       \bottomrule
    \end{tabular}
    }
        \caption*{List}
    \end{minipage}
    \hfill
    \begin{minipage}{0.32\textwidth}
        \centering
        \resizebox{\textwidth}{!}{
        \begin{tabular}{ccccccccc}    \toprule
       $|S|$/$10^5$ & 1 &3  &5  &10  & 20 & 30 & 50 & 100     \\ 
       \midrule
       $TV$  & 0.542 & 0.336 & 0.248 & 0.178 & 0.126 & 0.103 & 0.080 & 0.056    \\ 
       $Flat_2$  & 0.359 & 0.242 & 0.188 & 0.135 & 0.096 & 0.078 &0.060 & 0.043   \\ 
       $Flat_{20}$ & 0.525 & 0.306 & 0.202 & 0.106 & 0.060 & 0.045 &0.031 & 0.020   \\ 
       \bottomrule
    \end{tabular}
    }
        \caption*{PCFG}
    \end{minipage}
    \hfill
    \begin{minipage}{0.32\textwidth}
        \centering
        \resizebox{\textwidth}{!}{
        \begin{tabular}{ccccccccc}    \toprule
       $|S|$/$10^5$ & 1 &3  &5  &10  & 20 & 30 & 50 & 100     \\ 
       \midrule
       $TV$  & 0.673 & 0.471 & 0.351 & 0.263 & 0.177 & 0.146 & 0.125 & 0.093    \\
       $Flat_2$  & 0.371 & 0.275 & 0.224 & 0.162 & 0.118 & 0.100 &0.081 & 0.063   \\ 
       $Flat_{20}$ & 0.497 & 0.317 & 0.234 & 0.138 & 0.082 & 0.063 &0.047 & 0.033   \\ 
       \bottomrule
    \end{tabular}
    }
        \caption*{Markov}
    \end{minipage}
\end{table*}

\begin{table*}[htbp]
    \centering
    \caption{Sample Complexity of PPMs (duowan, $|D|=49.8\times 10^5$)}
    \begin{minipage}{0.32\textwidth}
        \centering
        \resizebox{\textwidth}{!}{
        \begin{tabular}{ccccccccc}    
    \toprule
       $|S|$/$10^5$ & 5 &10  &30  &50  & 100 & 300 & 500 & 1000     \\ 
       \midrule
       $TV$  &0.676  & 0.585 & 0.374 & 0.258 & 0.189 & 0.112 & 0.087 & 0.062    \\ 
       $Flat_2$  &0.421  & 0.375 & 0.261 & 0.202 & 0.147 & 0.086 & 0.067 & 0.047   \\ 
       $Flat_{20}$ &0.671  & 0.579 & 0.361 & 0.245 & 0.124 & 0.051 & 0.036 & 0.023   \\ 
    \bottomrule
    \end{tabular}
        }
        \caption*{List}
    \end{minipage}
    \hfill
    \begin{minipage}{0.32\textwidth}
        \centering
        \resizebox{\textwidth}{!}{
        \begin{tabular}{ccccccccc}    
    \toprule
       $|S|$/$10^5$ & 5 &10  &30  &50  & 100 & 300 & 500 & 1000      \\ 
    \midrule
       $TV$  & 0.569 & 0.481 & 0.304 & 0.214 & 0.155 & 0.091 &0.071 & 0.050   \\ 
       $Flat_2$  & 0.385 & 0.336 & 0.227 & 0.173 & 0.125 & 0.073 &0.057 & 0.040    \\ 
       $Flat_{20}$ & 0.557 & 0.468 & 0.282 & 0.191 & 0.102 & 0.044 &0.031 & 0.020    \\ 
    \bottomrule
    \end{tabular}
    }
        \caption*{PCFG}
    \end{minipage}
    \hfill
    \begin{minipage}{0.32\textwidth}
        \centering
        \resizebox{\textwidth}{!}{
        \begin{tabular}{ccccccccc}    
    \toprule
       $|S|$/$10^5$ & 5 &10  &30  &50  & 100 & 300 & 500 & 1000      \\ 
    \midrule
       $TV$  &0.247  & 0.203 & 0.138 & 0.115 & 0.091 & 0.068 &0.061 & 0.056     \\
       $Flat_2$  &0.198  & 0.162 & 0.111 & 0.091 & 0.070 & 0.051 &0.047 & 0.041   \\ 
       $Flat_{20}$ &0.202  & 0.158 & 0.096 & 0.072 & 0.048 & 0.029 &0.024 & 0.019   \\ 
    \bottomrule
    \end{tabular}
    }
        \caption*{Markov}
    \end{minipage}
\end{table*}

\begin{figure*}[htbp]
    \centering
    \captionsetup[subfigure]{labelformat=empty}
    \begin{minipage}{0.32\textwidth}
        \centering
        \includegraphics[width=\textwidth]{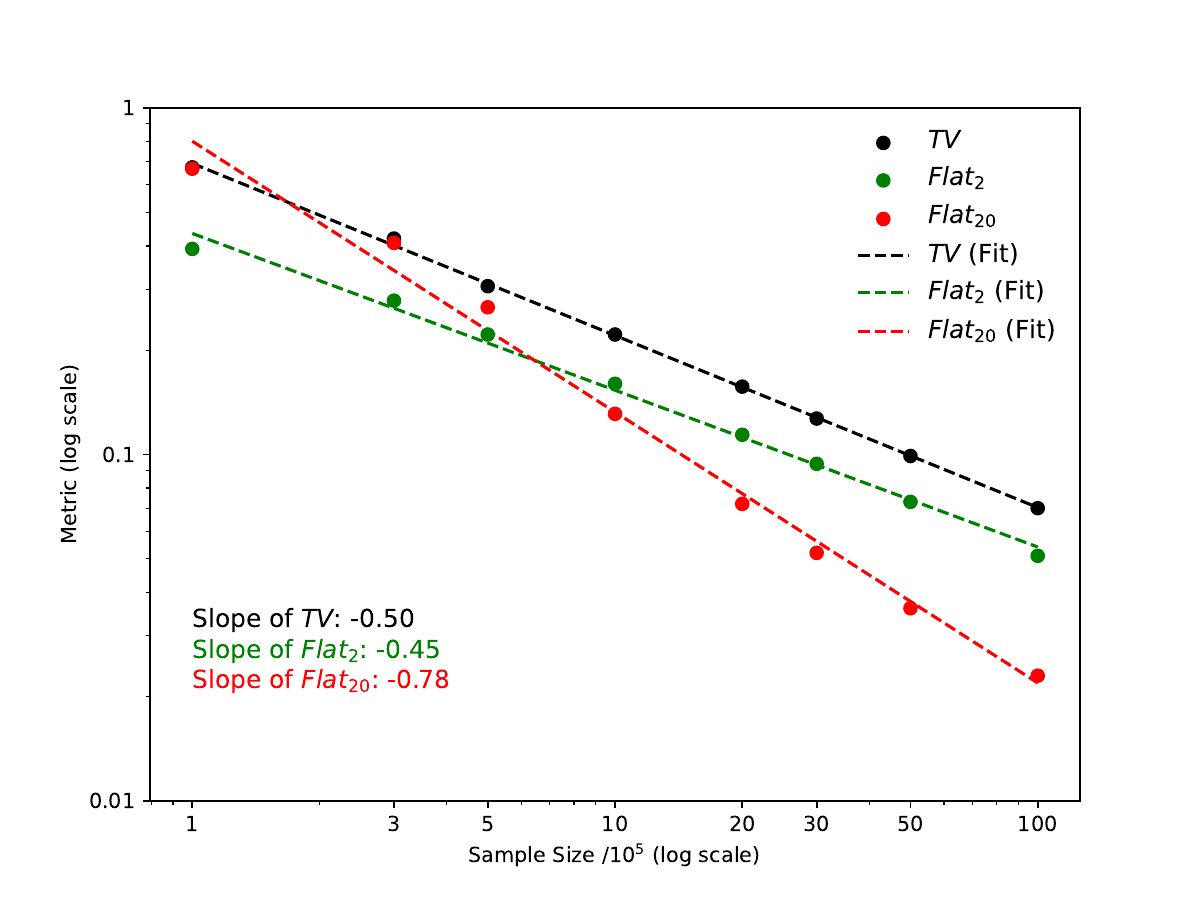}
        \caption*{List}
    \end{minipage}
    \hfill
    \begin{minipage}{0.32\textwidth}
        \centering
        \includegraphics[width=\textwidth]{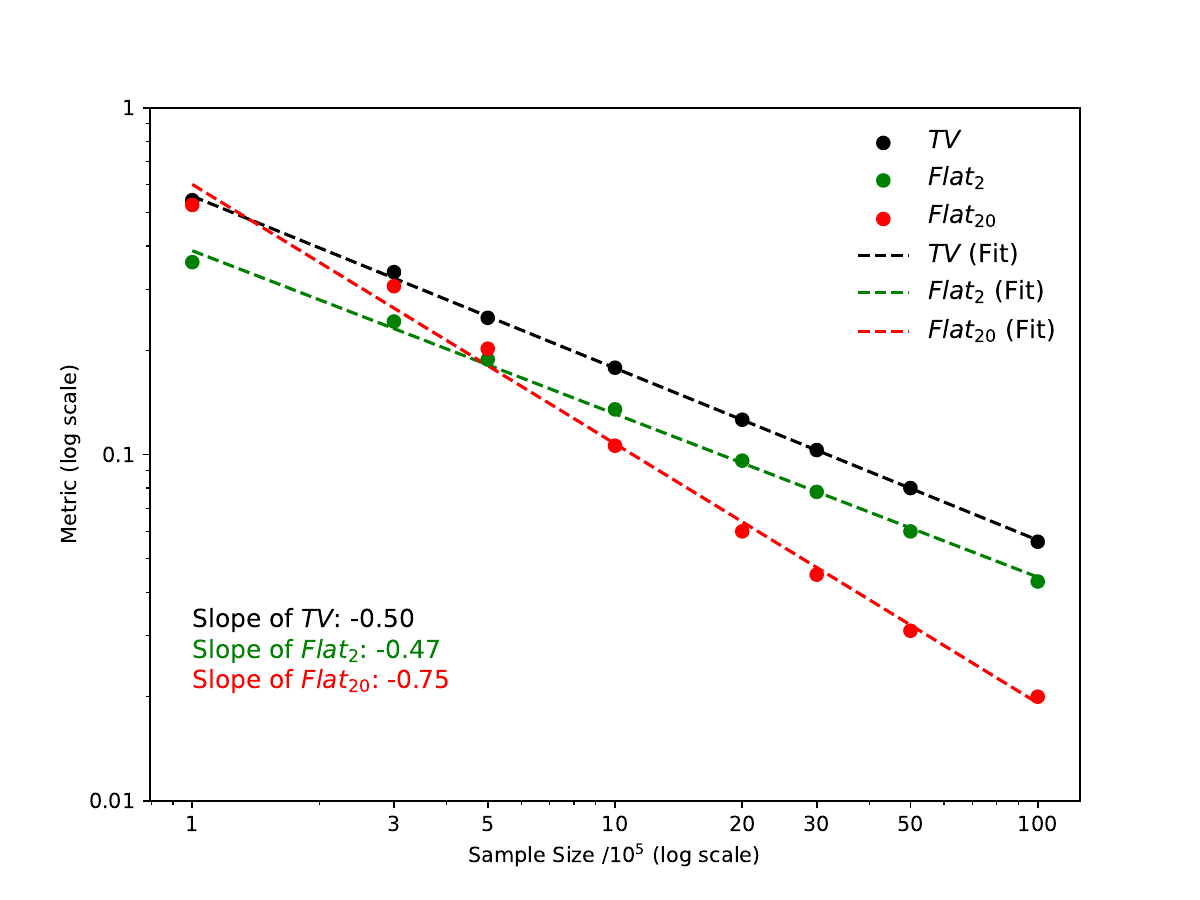}
        \caption*{PCFG}
    \end{minipage}
    \hfill
    \begin{minipage}{0.32\textwidth}
        \centering
        \includegraphics[width=\textwidth]{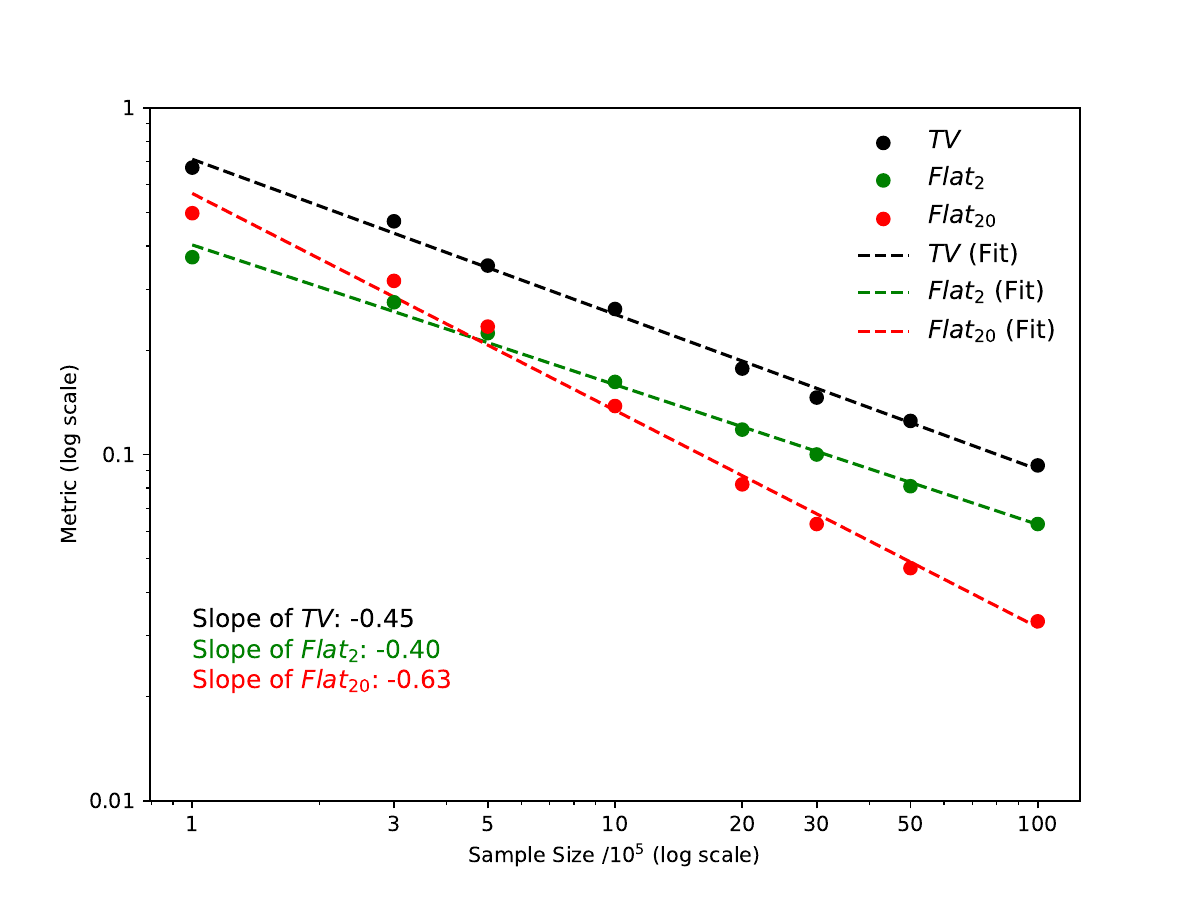}
        \caption*{Markov}
    \end{minipage}
    \caption{Log-Log Plot of Sample Size vs Metric with Linear Regression (Yahoo, $|D|=4.4\times 10^5$)}
    \label{fig_yahoo}
\end{figure*}

\begin{figure*}[htbp]
    \centering
    \captionsetup[subfigure]{labelformat=empty}
    \begin{minipage}{0.32\textwidth}
        \centering
        \includegraphics[width=\textwidth]{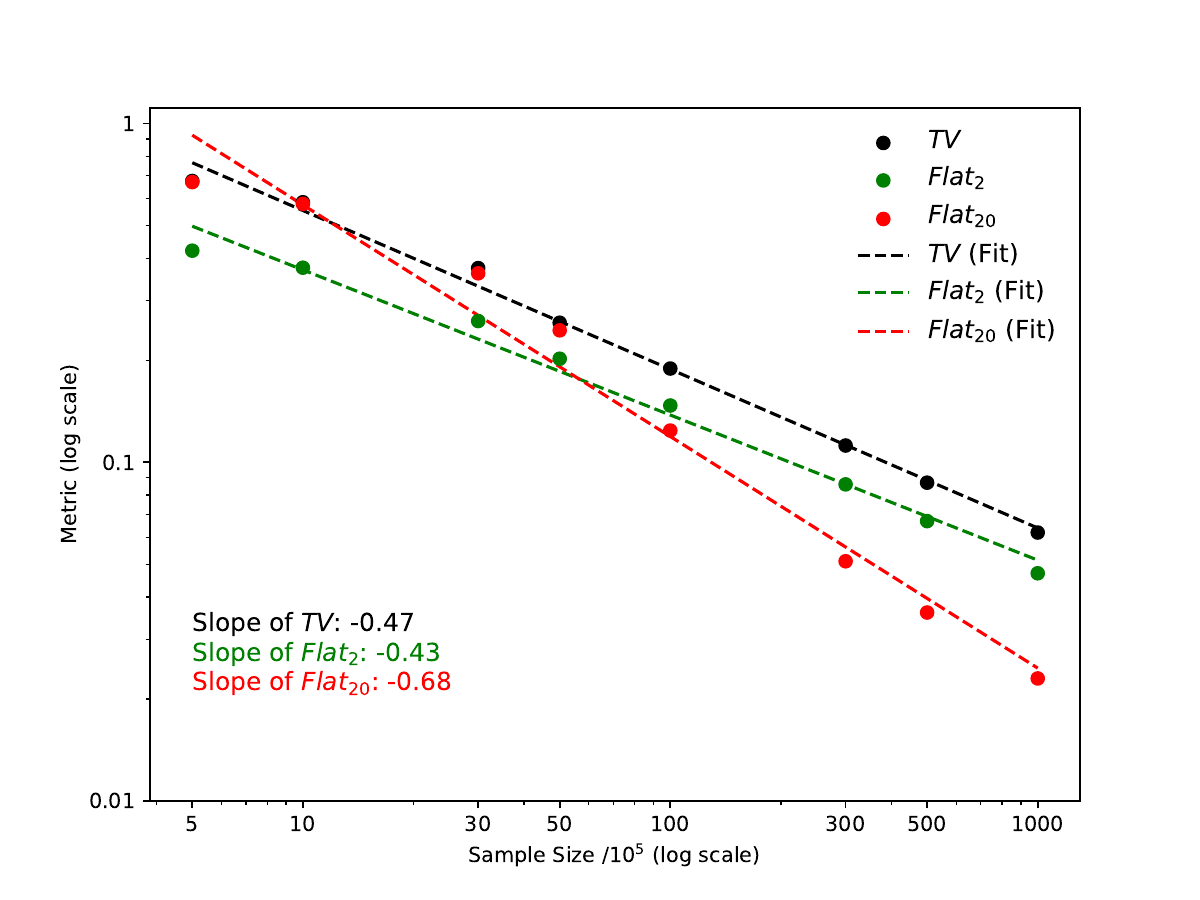}
        \caption*{List}
    \end{minipage}
    \hfill
    \begin{minipage}{0.32\textwidth}
        \centering
        \includegraphics[width=\textwidth]{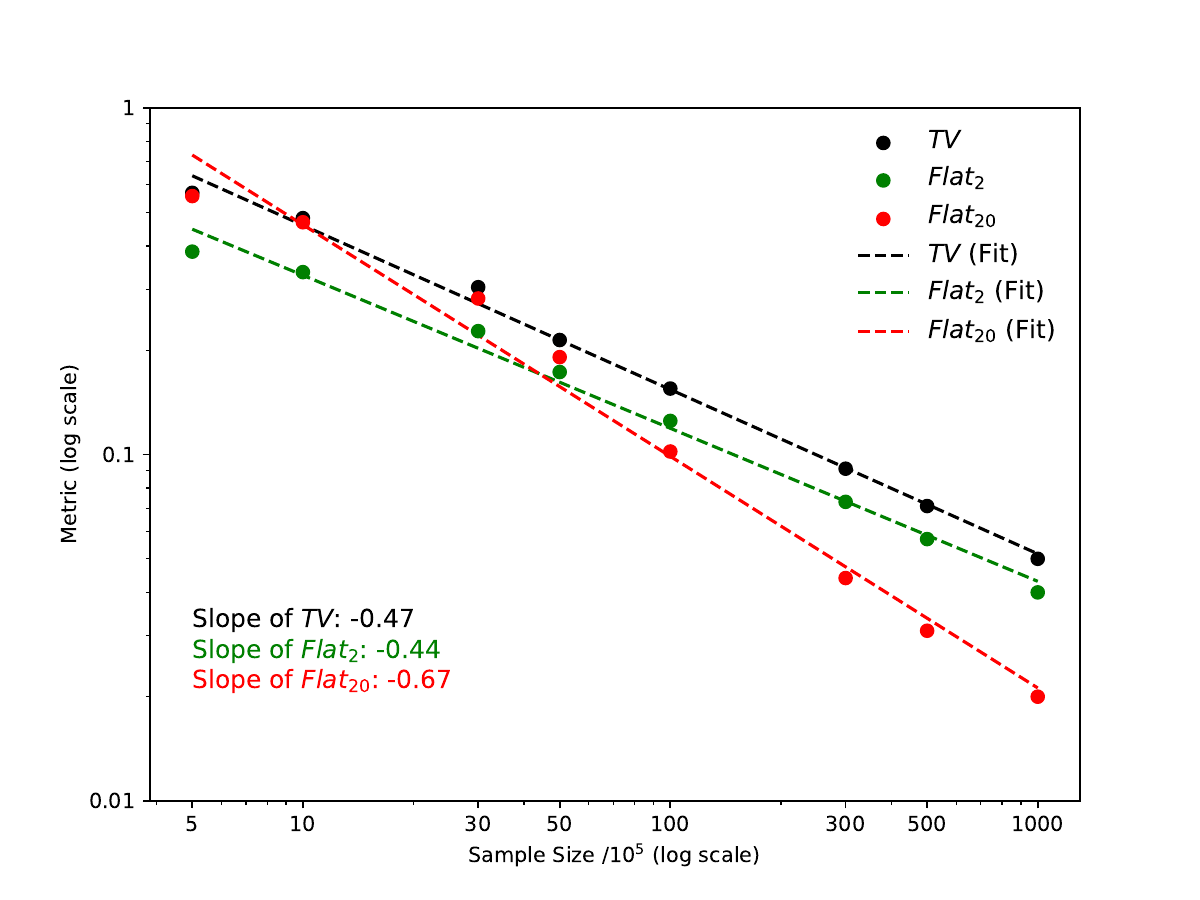}
        \caption*{PCFG}
    \end{minipage}
    \hfill
    \begin{minipage}{0.32\textwidth}
        \centering
        \includegraphics[width=\textwidth]{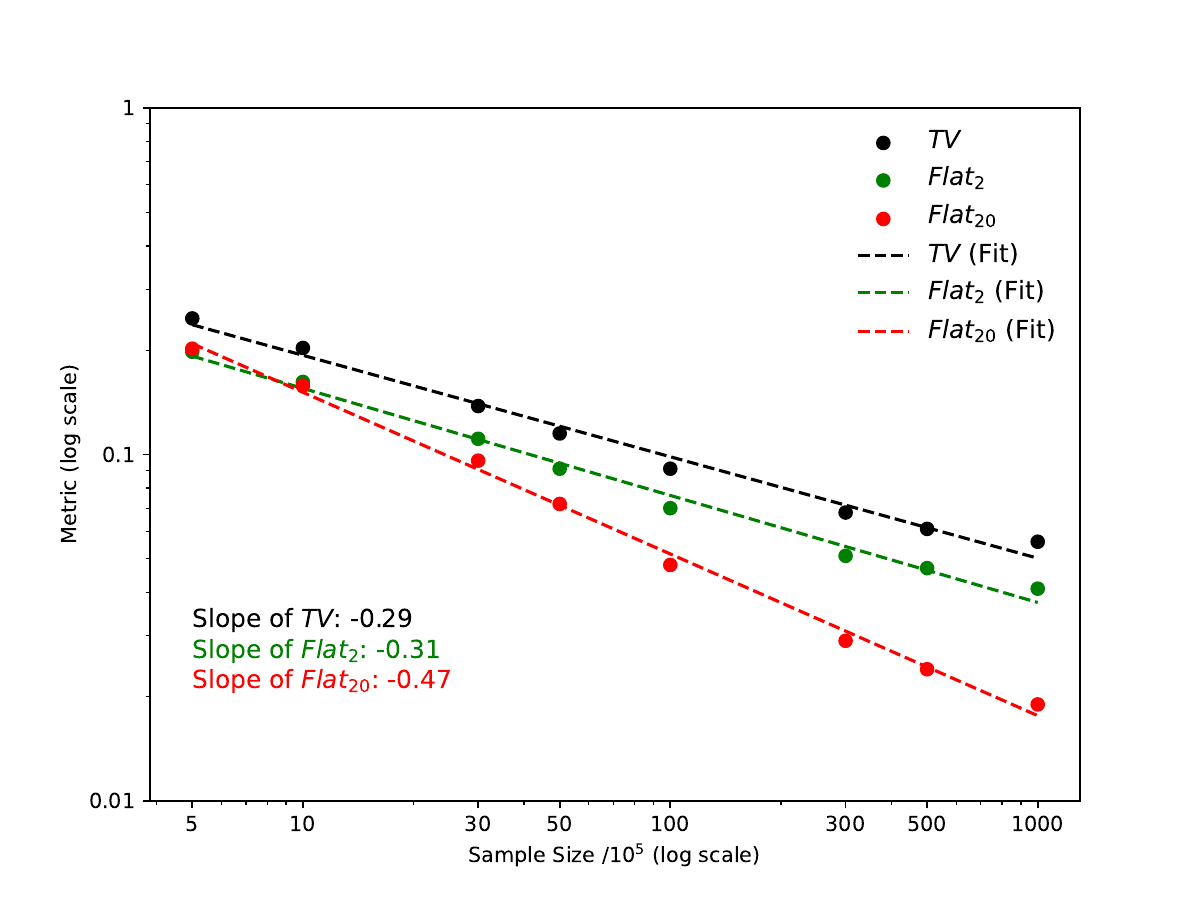}
        \caption*{Markov}
    \end{minipage}
    \caption{Log-Log Plot of Sample Size vs Metric with Linear Regression (Duowan, $|D|=49.8\times 10^5$)}
\end{figure*}

Experimental results show that to fit the target distribution derived from a small password dataset of 0.13 million entries, all models require samples on the order of millions in order to achieve a total variation distance between the model and the target distribution of less than 0.1 \footnote{When the total variation (TV) is set to 0.1, the upper bound of the attacker's first attack success rate is given by $0.1 + \frac{1}{k}$ (according to Theorem \ref{crypto-game}). In practical applications, the TV value of interest can be selected based on the desired level of security.
}. When the size of the password dataset reaches five million, the required sample size approaches thirty million.

Additionally, we found that both the total variation distance and the $Flat_2$ of the List model and the PCFG model are approximately proportional to $\frac{1}{\sqrt{|S|}}$, which is guaranteed by the central limit theorem. This results in a straight line on a double-logarithmic scale when plotting the data.

From the perspective of sample complexity, the Markov model exhibits distinct characteristics compared to the List and PCFG models. Specifically, the slope parameters in the log-log plot for the Markov model range between 0.3 and 0.45, whereas those for the other two models are centered around 0.5. Additionally, the training difficulty for the Markov model remains relatively consistent as the dataset size varies from $4.4 \times 10^5$ to $49.8 \times 10^5$. In contrast, the List and PCFG models demonstrate a significant dependency on dataset size.

\section{Discussion and Limitations}
\subsection{Reflect on Our Theoretical Lens}
From a game-theoretic perspective, the honeyword problem can be restated as follows: Suppose the true password distribution of users is denoted by $P$, and a website, leveraging its knowledge (e.g., public datasets or registered user passwords), determines a honeyword distribution $Q$. The attacker, based on their own knowledge, forms estimates $P'$ and $Q'$ for the user’s true password distribution and the honeyword distribution, respectively. If we assume the attacker has perfect knowledge of the website's honeyword distribution, i.e., $Q' = Q$, then the attacker's capability depends on the discrepancy between $P'$ and $P$. The closer $P'$ is to $P$, the stronger the attacker's ability to distinguish honeywords from real passwords, but it is still bounded by the information-theoretic limit $\epsilon_k(P, Q)$.

However, achieving this information-theoretic security is highly challenging. On one hand, learning the password distribution requires a substantial sample size (sample complexity); on the other hand, $P$ may not be unique—it can vary across websites and communities, change over time, or even be a conditional probability distribution dependent on user-specific information. These complexities motivate experimental studies across various scenarios \cite{21,40}. Previous studies have conducted experiments to evaluate the attack capabilities of an adversary with knowledge of $P'$ and $Q'=Q$. Their conclusions depend highly on experimental settings, datasets, and assumptions about the attacker's capabilities. In contrast, this study focuses on the strongest possible attacker and does not depend on $P'$ and $Q'$.

Moreover, our study on sample complexity provides an analysis of the attacker's ability to estimate $Q$ when $Q$ is not directly available. In practice, the attacker may not have direct access to the website's honeyword generation model but can attempt to estimate or reconstruct it using samples generated by the model. The number of samples required for this estimation corresponds to the sample complexity of the password probability model.

\subsection{Application of Our Theoretical Results}
Our theoretical results have several direct applications:

For given probability distributions $P$ and $Q$, we can theoretically compute the flatness and success-number curves of the strongest attacker. Previously, these curves could only be estimated through Monte Carlo experiments.

From the theoretical analysis of flatness (eq. (\ref{f4})), we derive the following asymptotic relationship: $\epsilon_k(1) = \Theta(\frac{M}{k}) + f(+\infty)$. This relationship provides practical guidance for selecting the parameter $k$. Specifically, if $f(+\infty) = 0$ and $M$ is not large, doubling $k$ would nearly halve $\epsilon_k(1)$ (see Example \ref{exmp1}).

If the honeyword distribution follows a uniform distribution, our framework allows for precise computation and analysis, which is shown in Example \ref{exmp1} and \ref{exmp2}. On one hand, using a uniform honeyword distribution does not require additional knowledge about the true password distribution. On the other hand, it exhibits a certain level of robustness when the user's password distribution $P$ changes, either over time or due to the attacker's knowledge of user-specific information, making it a conditional distribution. Our theoretical analysis provides a deeper understanding of the uniform honeyword distribution.

\subsection{Limitation of flatness calculation}

The limitation of our calculation of the flatness is that the actual situation requires $k$ sweetwords to be different. Fortunately, further analysis indicates that this restriction has only a negligible impact on our results.

In the following, we estimate the probability of the condition when at least two sweetwords are sampled the same (denoted as Event $A$). Define that Event $B$ is that a honeyword and a password are sampled the same, and Event $C$ is that there are two honeywords sampled the same. Using the union bound, we can get:
\begin{align*}
    \Pr&[A] \le \Pr[B]+\Pr[C]\\
    &\le \sum_{pw}P(pw)\{1-[1-Q(pw)]^{k-1}\} +\tbinom{k-1}{2} \sum_{pw}Q^2(pw) .
\end{align*}
We can conclude that the flatness under the requirement of different sweetwords is less than the flatness calculated in Section \ref{sec_flat} plus $\Pr[A]$.
Fortunately, $\Pr[A]$ is small in most cases so our flatness calculation is satisfactory.
For example, if $Q$ is a uniform distribution over password space, $Q(pw)=\frac{1}{n},pw \in \mathcal{PW}$. In this case, $\Pr[A]\le 1-(1-\frac{1}{n})^{k-1}+\frac{k^2}{2n}\le \frac{k^2+2k-2}{2n}$. Consider that the typical size of password space is larger than $10^6$, so this probability is smaller than $5\times 10^{-3}$ when $k\le 100$. 

\subsection{Limitation of success-number calculation}
The success-number function is determined based on the probability distribution of $w$ (i.e., $a(x)$), but deriving a neat mathematical expression for $a(x)$ is challenging, even in simple cases where $P$ and $Q$ follow uniform or Zipf distributions. However, this does not prevent numerical computation using the formula derived from our theorem \ref{therm_lambda}. Through experiments, we find that the numerical results obtained from the formula closely match the empirical curves of the strongest attacker.

\section{Conclusion}
Our work delves deeply into the mathematical foundation for the information-theoretic security analysis of Honeyword systems, providing, for the first time, mathematical formulations for the two key security metrics of the strongest attacker in Honeyword schemes. This theoretical contribution reveals the functional relationship between Honeyword security and system parameters $k$ and $T_2$. An unexpected theoretical result is our proof that Flatness and total variation are bounded by constant factors of each other, bridging the theoretical study of Honeyword systems with classical information theory.

To investigate the security of existing Honeyword generation methods based on password probability models, we conducted, for the first time, an exploration of the sample complexity of three classical PPMs. In this process, we proposed polynomial-time approximation algorithms for computing the total variation distance for PCFGs and higher-order Markov models. Our experimental results provide insights into the sample size required for PPM training in practice, as well as the fitting error and convergence rate.
We believe that this work will contribute to advancing research on Honeyword systems and password probability models.

\bibliographystyle{IEEEtran}
\bibliography{sample-base}
\newpage

{\appendices

\section{Missed Proof}
\subsection{Proof of Theorem \ref{therm3}}\label{prooftherm3}
In the discrete case, there is a special scenario: a sweetword list contains $i$ sweetwords with identical $P/Q$ values. In this scenario, one optimal strategy for guessing is to randomly select one of the $i$ sweetwords with identical $P/Q$ values (as outlined by the Bayesian formula (\ref{f1})). Although any selection among these sweetwords is equally optimal, choosing one at random facilitates the subsequent mathematical analysis.
\begin{proof}
    It can be calculated as follows:
\begin{align*}
\epsilon_k(1)&=\sum_{x}\sum_{i=1}^{k}\frac{1}{i}f(x)\tbinom{k-1}{i-1}G^{k-i}(x^-)g^{i-1}(x)\\
&=\sum_{x}\sum_{i=1}^{k}\frac{1}{k}f(x)\tbinom{k}{i}G^{k-i}(x^-)g^{i-1}(x)\\
&=\sum_{x}\sum_{i=1}^{k}\frac{1}{k}xg(x)\tbinom{k}{i}G^{k-i}(x^-)g^{i-1}(x)\\
&=\sum_{x}\frac{1}{k}x[(G(x^-)+g(x))^k-G^k(x^-)]\\
&=\sum_{x}\frac{1}{k}x[G^{k}(x)-G^{k}(x^-)].
\end{align*}
The key steps use a combinatorial identity and the binomial theorem.
\end{proof}

\subsection{Calculation of $f(+\infty)$ of the List Model}\label{appendix0}

Denote the $i$-th largest probability in a Zipf distribution as $p_i$. With the help of the Taylor series $e^{-x}=1-x+\frac{x^2}{2}-\frac{x^3}{6}+\cdots$, we can do the following calculation:
\begin{align*}
    \mathbb{E}[f(+\infty)]&=\sum_{i=1}^{n}p_i\mathbb{E}[1_{pw_i  \notin S}]\\
    &=\sum_{i=1}^{n}p_i(1-p_i)^{\vert S \vert}\\
    &\approx \sum_{i=1}^{n}p_i\cdot e^{-p_i\vert S \vert }\\
    &=\sum_{i=1}^{n}p_i\cdot (1-p_i\vert S \vert +\frac{1}{2!}(p_i\vert S \vert)^2-\cdots)\\
    &=\sum_{i=1}^{n} \sum_{j=0}^{+\infty} (-1)^{j}\frac{1}{j!} p_i^{j+1} \vert S \vert^j\\
    &=\sum_{j=0}^{+\infty} (-1)^{j}\frac{1}{j!}\vert S \vert^j \sum_{i=1}^{n} p_i^{j+1} .
\end{align*}
Since $p_i=\frac{i^{-\alpha}}{A}$ where $A=\sum_{k=1}^{n}k^{-\alpha}$, it can be calculated that $\sum_{i=1}^{n}p_i=1$, $\sum_{i=1}^{n}p_i^j=\frac{1}{A^j}\sum_{i=1}^{n}i^{-j\cdot\alpha} \approx \frac{1}{A^j} \zeta(j\cdot\alpha)$ when $j\cdot\alpha >1$. $\zeta$ is the Riemann zeta function. Consider $\alpha=0.7$ \cite{21}, we get 
$$
\mathbb{E}[f(+\infty)]\approx 1+\sum_{j=1}^{+\infty} (-1)^{j}\frac{1}{j!} \frac{\vert S \vert^j}{A^{j+1}} \zeta(0.7(j+1)).
$$
Since the Riemann zeta function is monotonically decreasing in $(1,+\infty)$, an estimate can be obtained by scaling this term to a constant.

\subsection{Proof of Theorem \ref{crypto-game} }\label{appendix2}
We will prove the two inequalities separately.
\begin{proof}
We construct an attacker $\mathcal{A}$ as follows: The attacker randomly selects a random index $r \gets \{1, 2, \cdots, k \}$. If $P(sw_r) \ge Q(sw_r)$, the attacker guesses $r$ as the index of the real password. Otherwise, the attacker randomly selects another index (other than $r$) as the guess.

Next, we calculate the attack success probability of $\mathcal{A}$. Denote the index of the real password is $y$. When $r = y$, the attacker must submit $r$ as the guess to succeed in the attack. According to $\mathcal{A}$'s attack algorithm, this requires the true passphrase $sw_r$ to satisfy $P(sw_r) > Q(sw_r)$. Since $sw_r \sim P$, the probability of this event is given by $\sum_{x:P(x)>Q(x)}P(x)$. 
When $r \neq y$, the attacker must submit a random guess other than $r$ in order to succeed with a probability of $\frac{1}{k-1}$. According to $\mathcal{A}$'s attack algorithm, this requires the false passphrase $sw_r$ to satisfy $P(sw_r) < Q(sw_r)$. Since in this case $sw_r \sim Q$, the probability of this event is given by $\sum_{x:P(x)<Q(x)}Q(x)$. 
Combining the above analysis, we can calculate the overall attack success probability as follows:
\begin{align*}
\epsilon_k^{\mathcal{A}}(1)&=\Pr[\mathcal{A}\text{ wins}| y=r]\Pr[y=r]\\
&\ \ \ \ \ \ +\Pr[\mathcal{A}\text{ wins}| y\neq r]\Pr[y\neq r]\\
&=\frac{1}{k}\Pr[\mathcal{A} \text{ wins}| y=r]+\frac{k-1}{k}\Pr[\mathcal{A} wins| y\neq r]\\
&=\frac{1}{k} \sum_{x:P(x)>Q(x)}P(x) +\\
&\ \ \ \ \ \ \frac{k-1}{k}\cdot \left( \frac{1}{k-1} \sum_{x:P(x)<Q(x)}Q(x) \right) \\
&=\frac{1}{k} \sum_{x:P(x)>Q(x)}P(x) +\frac{1}{k}\sum_{x:P(x)<Q(x)}Q(x) \\
&=\frac{1}{k} \sum_{x:P(x)>Q(x)}[\frac{P(x)+Q(x)}{2}+\frac{P(x)-Q(x)}{2}]\\
&\ \ \ \ \ +\frac{1}{k}\sum_{x:P(x)<Q(x)}[\frac{P(x)+Q(x)}{2}+\frac{Q(x)-P(x)}{2}] \\
&=\frac{1}{k}\sum_{x}[\frac{P(x)+Q(x)}{2}+\frac{|P(x)-Q(x)|}{2}]\\
&=\frac{1}{k}+\frac{1}{k}\Delta_{TV}(P,Q) \qedhere
\end{align*}

\end{proof}

Next, we prove the other side of the inequality.
\begin{proof}
Recall the definitions of $f$ and $g$ (\ref{def_f}, \ref{def_g}). A key observation is that \[\Delta_{TV}(P,Q)=\frac{1}{2}\int_{0}^{M}|f(x)-g(x)|\mathrm{d}x +\frac{1}{2}f(+\infty).\]

With the help of the property (\ref{prop1}), it can be calculated as:
\begin{align*}
    &\Delta_{TV} (P,Q)=\frac{1}{2}\int_{0}^{M}|f(x)-g(x)|\mathrm{d}x +\frac{1}{2}f(+\infty)\\
    &=\frac{1}{2}\int_{0}^{M}|x-1|g(x)\mathrm{d}x +\frac{1}{2}f(+\infty)\\
    &=\frac{1}{2}\int_{0}^{1}(1-x)g(x)\mathrm{d}x +\frac{1}{2}\int_{1}^{M}(x-1)g(x)\mathrm{d}x +\frac{1}{2}f(+\infty)\\
    &=\frac{1}{2}[(1-x)G(x)|_0^1+\int_0^1G(x)\mathrm{d}x+(x-1)G(x)|_1^M\\
    & \ \ \ \ \      -\int_1^M G(x)\mathrm{d}x +f(+\infty)]\\
    &=\frac{1}{2}\left[2\int_0^1G(x)\mathrm{d}x+M-1-\int_0^M G(x)\mathrm{d}x +f(+\infty)\right]\\
    &=\int_0^1G(x)\mathrm{d}x.
\end{align*}

This leads to $\int_0^1G(x)\mathrm{d}x = \Delta(P,Q)$. Due to the definition of Total Variance, we have $0 \le f(+\infty)\le \Delta(P,Q)$. Thus $\int_1^M G(x)\mathrm{d}x=\int_0^M G(x)\mathrm{d}x-\int_0^1 G(x)\mathrm{d}x=M-1+f(+\infty)-\Delta_{TV}(P,Q)$.

Recall the formula (\ref{f4}), 
\begin{align*}
    Flat_k&(P||Q) =\frac{1}{k}(M-\int_0^M G^k(x)\mathrm{d}x)+f(+\infty)-\frac{1}{k}\\
    &\le \frac{1}{k}(M-1-\int_1^M G^k(x)\mathrm{d}x)+f(+\infty) \\
    &= \frac{1}{k}(M-1)\left[1-\left( \int_1^M \frac{1}{M-1} G^k(x)\mathrm{d}x\right)\right]+f(+\infty)\\
    &\le \frac{1}{k}(M-1)\left[1- \left( \int_1^M \frac{1}{M-1} G(x)\mathrm{d}x\right)^k\right]+f(+\infty)\\
    &= \frac{1}{k}(M-1)\left[1-\left( \frac{(M-1+f(+\infty)-\Delta(P,Q))}{(M-1)}\right)^k \right]\\
    & \ \ \ \ \ \ \ \ \ +f(+\infty)\\
    &\le \frac{1}{k}(M-1)\left[1-\left(1+k\frac{f(\infty)-\Delta(P,Q)}{M-1} \right)\right]+f(+\infty)\\
    &=\Delta_{TV}(P,Q)
\end{align*}

The core steps in the above process use Jensen's inequality (the convexity of $x^k$) and the Bernoulli inequality $(1+x)^n \geq 1 + nx, \forall x \geq -1$.   
\end{proof}

\section{Proof of Theorem \ref{therm_lambda} }\label{appendix_sn}

\subsection{A Toy Proof on $\lambda(1)$}
Here we give a toy proof of $\lambda_U(1)$ and leave the full proof of $\lambda_U(i), 1\le i \le U$ in the next section.
\begin{proof}[Proof (toy)]
Recall the definition of $\lambda_U(1)$. It is the expectation of the success-number of the strongest attacker before its first failure. Order all $w_i, 1\le i\le U$ into $W_1 \ge W_2 \ge \cdots \ge W_U$. In this case, the expectation of the success-number can be calculated as:
\begin{align*}
    &W_1(1-W_2)+2W_1W_2(1-W_3)+\cdots\\
    &\ \ \ \ \ \ \  +(U-1)W_1\cdots W_{U-1}(1-W_U) +U\cdot W_1W_2\cdots W_U\\
    &=W_1+W_1W_2+\cdots +W_1W_2\cdots W_U\\
    &=\sum_{i=1}^{U}\prod_{j=1}^{i}W_j.
\end{align*}
Take expectation on both sides so we can get the expression of $\lambda_U(1)$:
\begin{align*}
    \lambda_U(1)&=\mathbb{E}[\sum_{i=1}^{U}\prod_{j=1}^{i}W_j]=\sum_{i=1}^{U}\mathbb{E}[\prod_{j=1}^{i}W_j],
\end{align*}
where $W_j$ is the $j$-th largest $w$ of all $U$ Users.

The next task is to calculate $\mathbb{E}[\prod_{j=1}^{i}W_j]$. Fortunately, a conditional expectation of it is feasible to calculate:
\begin{align*}
    &\mathbb{E}[\prod_{j=1}^{i}W_j\vert W_i=t]\\
    &=\tbinom{U}{1}\tbinom{U-1}{i-1}\int_t^1\int_t^1\cdots \int_t^1 x_1x_2\cdots x_{i-1}\cdot t \cdot 
    a(x_1)\cdot \\
    &\ \ \ a(x_2)\cdots a(x_{i-1})\left(\int_0^t a(x)\right)^{U-i} \mathrm{d}x_1\mathrm{d}x_2\cdots \mathrm{d}x_{i-1}\\
    &=t\tbinom{U}{1}\tbinom{U-1}{i-1}\left(\int_t^1 xa(x)\mathrm{d}x\right)^{i-1} \left(\int_0^t a(x)\mathrm{d}x\right)^{U-i}.
\end{align*}

The following process can help understand this conditional expectation: First, we select one of the $U$ users as the $t$-th largest $w$ which is $t$. Next, we select $(i-1)$ users whose $w$ are $x_1,x_2,\cdots,x_{i-1} \ge t$. The rest users' $w$ are less than $t$. In this case, $\prod_{j=1}^{i}W_j=x_1x_2\cdots x_{i-1}\cdot t$ and the probability can be calculated with the distribution $a(w)$.
With this conditional expectation, we can calculate the sum:
\begin{align*}
    &\sum_{i=1}^{U}\mathbb{E}[\prod_{j=1}^{i}W_j\vert W_i=t]\\
    &=\sum_{i=1}^{U}t\tbinom{U}{1}\tbinom{U-1}{i-1}\left(\int_t^1 xa(x)\mathrm{d}x\right)^{i-1} \left(\int_0^t a(x)\mathrm{d}x\right)^{U-i}\\
    &=tU\left(\int_0^t a(x)\mathrm{d}x+\int_t^1 xa(x)\mathrm{d}x\right)^{U-1}\\
    &=tU(\psi(t))^{U-1},
\end{align*}
The key step in the above calculation uses the inverse form of the binomial theorem.
All are prepared so that we can reach $\lambda_U(1)$:
\begin{align*}
    \lambda_U(1)
    &=U\int_{\frac{1}{k}}^1 ta(t)\left(\psi(t)\right)^{U-1}\mathrm{d}t .\qedhere
\end{align*}
\end{proof}

The above calculation process is not rigorous, but it is helpful to understand the key ideas. The imprecision lies in that it is meaningless to discuss the probability of a continuous random variable taking a value at a certain point. This loophole can be solved by considering the interval ($\frac{1}{k} \le W_i\le t$), but the calculation will be more cumbersome.

It needs more technology to get $\lambda_U(i), i > 1$. The complete proof is in Appendix \ref{appendix1}.

\subsection{Full Proof of $\lambda_U(i)$}\label{appendix1}

Here we give out another method to calculate $\lambda_U(1)$. This method is convenient to be generalized to get all $\lambda_U(i)$.

Consider two random processes: the first is to get $U$ samples $\{w_i\}_{i \in \{1,2,...,U\}}$ from the distribution of $w$, whose probability distribution function is $a(x),0\le x \le 1$; the second is to attack each account and get the attack results. Denote the success of attack as 1 and the failure as 0. Thus we get a $U$-bit string $b_1b_2...b_U \in \{0,1\}^U$.

Consider a random variable $A_i$ defined as:
$$
A_i(b_1b_2...b_n)=b_i\prod_{j \neq i}1_{\{w_j\le w_i\} \cup \{b_j=1\} }=b_i\prod_{j \neq i} b_j^{1_{\{w_j>w_i\}} } .
$$
When $A_i$ is equal to 1, it means that all accounts that guessed before $i$-th account are all cracked. So the $i$-th account is guessed successfully before the first failure. On the other hand, $A_i=0$ means the $i$-th account is guessed unsuccessfully or it is guessed after the first failure. By symmetry, all $A_i$ have the same expectation. So 
$$
\lambda_U(1)=\sum_{i=1}^U \mathbb{E}[A_i]=U\cdot\mathbb{E}[A_i] .
$$

Recall the definition of $v_t$:
$$ v_t(x)=x^{1_{\{x\ge t\}}}=\left\{
\begin{aligned}
&x, x \ge t \\
&1, x < t 
\end{aligned}\ \ \ \ .
\right.
$$

Now we calculate $\mathbb{E}[A_i]$:
\begin{align*}
    \mathbb{E}[A_i]&=\mathbb{E}_1\mathbb{E}_2[A_i]\\
    &=\mathbb{E}_1\mathbb{E}_2[b_i\prod_{j \neq i} b_j^{1_{\{w_j>w_i\}} }] \\
    &=\mathbb{E}_1[\mathbb{E}_2[b_i]\prod_{j \neq i} \mathbb{E}_2[b_j^{1_{\{w_j>w_i\}}} ] ] \\
    &=\mathbb{E}_1[ w_i \prod_{j \neq i} w_j^{1_{\{w_j>w_i\}} } ]\\
    &=\int_0^1 t [\prod_{j \neq i} \mathbb{E}_1v_t(w_j)]a(t)\mathrm{d}t\\
    &=\int_0^1 \left(\psi(t) \right)^{U-1}ta(t)\mathrm{d}t,\\
\end{align*}
where
$$
\psi(t) =\int_0^t xa(x)\mathrm{d}x +\int_t^1 a(x)\mathrm{d}x .
$$
Therefore, 
$$
\lambda_U(1)=U\cdot \mathbb{E}[A_i]=U\cdot \int_0^1 (\psi(t))^{U-1}ta(t)\mathrm{d}t .
$$

Analogously, consider a random variable defined below:
\begin{align*}
B_i^S(b_1&b_2...b_U)\\
&=b_i\prod_{j \notin S \cup \{i\} }1_{ \{w_j \le w_i \} \cup \{b_j=1\} } \prod_{j \in S}1_{ \{w_j>w_i\} \cap \{b_j=0\} } \\
&=b_i\prod_{j \notin S \cup \{i\} }b_j^{1_{ \{w_j > w_i \} } }\prod_{j \in S}(1-b_j^{1_{ \{w_j>w_i\} } } ) ,
\end{align*}
where $S=\{i_1,i_2,...,i_{m-1}\}$. When $B_i^S=1$, attacker fails in the guesses of accounts in $S$, and succeeds in guessing other accounts whose $w$ is larger than $w_i$. It means that the $i$-th account is successfully guessed between the $(m-1)$-th failure and the $m$-th failure. Thus, 
$$
\lambda_U(m)-\lambda_U(m-1)=\sum_{i=1}^{n}\sum_{S\subset (U)-\{i\},\vert S \vert =m-1} \mathbb{E}[B_i^S] .
$$
Similar to the above calculation:
\begin{align*}
    \mathbb{E}[B_i^S]&=\mathbb{E}_1[\mathbb{E}_2 B_i^S]\\
    &=\int_0^1 [\psi(t)]^{n-k}[1-\psi(t)]^{k-1} t\cdot a(t)\mathrm{d}t .
\end{align*}
So we arrive at the final answer:
\begin{align*}
    \lambda_U(m)- \lambda_U(m-1)=\sum_{i=1}^{n}\sum_{S\subset (U)-\{i\},\vert S \vert =m-1} \mathbb{E}[B_i^S] \\
    =U\tbinom{U-1}{m-1}\int_0^1 [\psi(t)]^{n-k}[1-\psi(t)]^{k-1} t\cdot a(t)\mathrm{d}t ,
\end{align*}
where $\tbinom{U-1}{m-1}$ is the numbers of $S$.

To sum up, the formula of $\lambda_U(i)$ is :
\begin{align*}
    \lambda(i)=U\sum_{j=1}^{i}\int_{\frac{1}{k}}^{1}t\cdot a(t)\tbinom{U-1}{j-1}(\psi(t))^{U-j}(1-\psi(t))^{j-1}\mathrm{d}t .
\end{align*}

\section{Convolution property of success-number}\label{appendix_convolution}
\begin{proof}
The key step is to replace the integral variable. Let $\phi(x)$ be the inverse function of $\psi(t)$, and notice that $\frac{\mathrm{d} \psi(t) }{\mathrm{d} t}=(1-t)a(t)$, so
\begin{align*}
    \lambda_U&(i)=U\sum_{j=1}^{i}\int_{\frac{1}{k}}^{1}t\cdot a(t)\tbinom{U-1}{j-1}(\psi(t))^{U-j}(1-\mathbb{E}[v_t])^{j-1}\mathrm{d}t \\
    &=U\sum_{j=1}^{i}\int_{\frac{1}{k}}^{1} \frac{t}{1-t}\tbinom{U-1}{j-1}(\psi(t))^{U-j}(1-\mathbb{E}[v_t])^{j-1}\mathrm{d}(\psi(t) )\\
    &=U\sum_{j=1}^{i}\int_{\epsilon}^{1} \frac{\phi(x)}{1-\phi(x)} \tbinom{U-1}{j-1} x^{U-j}(1-x)^{j-1}\mathrm{d}x .\qedhere
\end{align*}

\end{proof}

\section{More Materials on Algorithms for Total Variation}
\subsection{Algorithm for Product Distribution}\label{appedix_tv1}
The mathematical techniques employed by Feng et al. in designing their approximation algorithm for the Total Variation distance of product distributions are, unexpectedly, consistent with those used in this work.
They define the likelihood ratio as a distribution, denoted by \((\mathbb{P}\| \mathbb{Q})\), such that 
\[
(\mathbb{P}\| \mathbb{Q})(r) := \Pr_{X \sim \mathbb{Q}} \left[ \frac{\mathbb{P}(X)}{\mathbb{Q}(X)} = r \right].
\]
Their definition is identical to the probability distribution $g$ defined in section \ref{sec_flat} of this work.

The likelihood ratio distribution (or “ratio” in short) \(\mathbb{R} = (\mathbb{P}\| \mathbb{Q})\) contains all the “useful” information about \((\mathbb{P}, \mathbb{Q})\). A key property is that \(\Delta_{\mathrm{TV}}(\mathbb{P}, \mathbb{Q}) = \mathbb{E}_{R \sim \mathbb{R}} \max(1 - R, 0)\).
This property can be expressed as an integral in the form $\Delta_{\mathrm{TV}}(P, Q) = \int_0^1 (1 - x) R(x) \mathrm{d}x$. Using the notation in this paper, it can be written as $\Delta_{\mathrm{TV}}(P, Q) = \int_0^1 (1 - x) g(x) \mathrm{d}x = \int_0^1 G(x) \mathrm{d}x$. A proof of this property is provided in Appendix \ref{appendix2}.

They denote the distance by \(\Delta_{\mathrm{TV}}(\mathbb{R})\). For the task of computing the total variation distance, it suffices to compute \(\mathbb{R} = (\mathbb{P}\| \mathbb{Q})\).
Given two product distributions \(\mathbb{P} = \mathbb{P}_1 \mathbb{P}_2 \ldots \mathbb{P}_n\) and \(\mathbb{Q} = \mathbb{Q}_1 \mathbb{Q}_2 \ldots \mathbb{Q}_n\), their ratio \((\mathbb{P}\| \mathbb{Q})\) can be expressed as:

\begin{align*}
 (\mathbb{P}\| \mathbb{Q})(r) &= \Pr_{(X_1, \ldots, X_n) \sim \mathbb{Q}} \left[ \frac{\mathbb{P}(X_1, \ldots, X_n)}{\mathbb{Q}(X_1, \ldots, X_n)} = r \right]\\
 &= \Pr_{X_1 \sim \mathbb{Q}_1, \, \dots, \, X_n \sim \mathbb{Q}_n} 
\bigg[
\underbrace{\frac{\mathbb{P}_1(X_1)}{\mathbb{Q}_1(X_1)}}_{R_1} 
\cdots 
\underbrace{\frac{\mathbb{P}_n(X_n)}{\mathbb{Q}_n(X_n)}}_{R_n} 
= r
\bigg] 
\end{align*}
which is the distribution of \(R_1 R_2 \ldots R_n\), where \(R_1, \ldots, R_n\) are independent and \(R_i \sim \mathbb{R}_i = (\mathbb{P}_i \| \mathbb{Q}_i)\).
This suggests a naïve algorithm to compute the total variation distance:
\begin{itemize}
    \item Compute \(\mathbb{R}_i = (\mathbb{P}_i \| \mathbb{Q}_i)\) for each \(i \in [n]\).
    \item Compute \(\mathbb{R} = \mathbb{R}_1 \cdot_{\mathrm{indp}} \mathbb{R}_2 \cdot_{\mathrm{indp}} \ldots \mathbb{R}_n\).
    \item Output \(\Delta_{\mathrm{TV}}(\mathbb{R})\).
\end{itemize}

Here, \(\mathbb{R}_1 \cdot_{\mathrm{indp}} \mathbb{R}_2\) denotes the distribution of \(R_1 R_2\) where \(R_1, R_2\) are independently sampled from \(\mathbb{R}_1, \mathbb{R}_2\). 
The naïve algorithm computes the exact value of the total variation distance, but has exponential time and space complexity because the support of \(\mathbb{R}\) can be exponentially large.

Their actual algorithm computes an approximation of \(\mathbb{R}\). The high-level framework of their algorithm looks as follows:
\begin{itemize}
    \item Compute \(\mathbb{R}_i = (\mathbb{P}_i \| \mathbb{Q}_i)\) for each \(i \in [n]\).
    \item Compute \(\mathbb{R}_1 \cdot_{\mathrm{indp}} \mathbb{R}_2\), then sparsify it as \(\widetilde{\mathbb{R}}_{1:2} \approx \mathbb{R}_1 \cdot_{\mathrm{indp}} \mathbb{R}_2\).
    \item Compute \(\widetilde{\mathbb{R}}_{1:2} \cdot_{\mathrm{indp}} \mathbb{R}_3\), then sparsify it as \(\widetilde{\mathbb{R}}_{1:3} \approx \widetilde{\mathbb{R}}_{1:2} \cdot_{\mathrm{indp}} \mathbb{R}_3\).

    $\vdots$
    \item Repeat this process iteratively until \(\widetilde{\mathbb{R}}_{1:n}\) is computed.
    \item Output \(\Delta_{\mathrm{TV}}(\mathbb{R}_{1:n})\) as an approximation of \(\Delta_{\mathrm{TV}}(\mathbb{P}, \mathbb{Q})\).
\end{itemize}

Sparsification reduces the size of the likelihood ratio distribution by grouping similar values while introducing a small, controlled error.
It consists of two steps: divide the range and merge probability masses.
First, it partitions the range of likelihood ratio values into intervals, using a logarithmic scale to ensure finer granularity for smaller values where the impact of the error is greater.
Within each interval, it combines all likelihood ratios into a single representative value, typically the weighted average of the likelihood ratios in that interval.
The sparsification reduces the support size of the likelihood ratio distribution from potentially exponential to a manageable size.

\subsection{Proof of Theorem \ref{therm_pcfg} }\label{appendix_tv2}

Reviewing the design of Algorithm 4, when constructing the auxiliary distributions $\mathbb{P}_0$ and $\mathbb{Q}_0$, multiplying $p$ and $q$ by the same scaling factor $\kappa \in [0,\frac{1}{\max(p,q)}]$ keeps the algorithm feasible. We will now prove that the error introduced by the algorithm is minimized when $\kappa = \frac{1}{\max(p, q)}$.

Define a distribution $\mathbb{P}_0$ and a distribution $\mathbb{Q}_0$:
$$\mathbb{P}_0(1)=\kappa \cdot p,\ \mathbb{P}_0(2)=1-\mathbb{P}_0(1),\ \mathbb{P}_0(3)=0.$$
$$\mathbb{Q}_0(1)=\kappa \cdot q,\ \mathbb{Q}_0(2)=0,\ \mathbb{Q}_0(3)=1-\mathbb{Q}_0(1).$$
When the scaling factor is $\kappa$, following the same computational logic. the total variation between $\mathbb{P}$ and $\mathbb{Q}$ is:



\begin{align*}
\Delta_{TV}(\mathbb{P},\mathbb{Q})&=\frac{1}{2}\sum_{pw \in \mathcal{PW}|_S}| \kappa\Pr_{M_1}[pw]-\kappa \Pr_{M_2}[pw]|\\
&\ \ \ \ \ \ +\frac{1}{2}[\mathbb{P}_0(2)+\mathbb{Q}_0(3)]\\
&= \frac{\kappa}{2}\sum_{pw \in \mathcal{PW}|_S}| \Pr_{M_1}[pw]-\Pr_{M_2}[pw]|+1-\kappa \frac{p+q}{2}
\end{align*}

Therefore, the total variation between the two PCFGs in the subspace $\mathcal{PW}|_S$ can be computed as:
$$ \frac{1}{\kappa}[\hat{\Delta}-(1-\kappa\cdot \frac{p+q}{2})]$$
where $\hat{\Delta}=TV4ProductDistritbuion(\mathbb{P},\mathbb{Q},\epsilon/2)$. 
Since the algorithm $TV4ProductDistribution$ introduces at most a relative error of $\frac{\epsilon}{2}$, the absolute error introduced in the above expression is at most:
\begin{align*}
\frac{1}{\kappa} \cdot \frac{\epsilon}{2} \Delta_{TV}(\mathbb{P},\mathbb{Q})&=\frac{1}{4}\sum_{pw \in \mathcal{PW}|_S}| \Pr_{M_1}[pw]-\Pr_{M_2}[pw]|+\frac{1}{2\kappa}\\
&\ \ \ \ \ -\frac{p+q}{4}
\end{align*}
The above expression decreases as $\kappa$ increases, and when $\kappa = \frac{1}{\max(p, q)}$, the error is minimized. The minimized error is 
\begin{align*}
\frac{1}{4}\sum_{pw \in \mathcal{PW}|_S}| \Pr_{M_1}[pw]-\Pr_{M_2}[pw]|+\frac{|p-q|}{4}
\end{align*}
Since our PCFG total variation estimation algorithm sums the results over all subspaces, the total absolute error does not exceed:
\begin{align*}
\sum_{S}&\frac{1}{4}\sum_{pw \in \mathcal{PW}|_S}| \Pr_{M_1}[pw]-\Pr_{M_2}[pw]|+\frac{|p-q|}{4}\\
&=\frac{1}{2}\Delta_{TV}(M_1,M_2)+\frac{1}{2}\Delta_{TV}(M_1[S],M_2[S])
\end{align*}
where $\Delta_{TV}(P_{M_1}[S], P_{M_2}[S])$ refers to the total variation (TV) distance between the syntax distributions of the two PCFGs, $M_1$ and $M_2$. 
By the post-processing property of total variation, we have:
$$\Delta_{TV}(M_1[S],M_2[S]) \le \Delta_{TV}(M_1,M_2).$$
Therefore, the total absolute error of our algorithm does not exceed $\epsilon \Delta_{TV}(M_1,M_2)$.

\section{More Experimental Results}\label{appendix_experiment}
We present the curves of the experimental results on the 12306 dataset in Figure \ref{fig_12306}.

\begin{figure*}[htbp]
    \centering
    \captionsetup[subfigure]{labelformat=empty}
    \begin{minipage}{0.32\textwidth}
        \centering
        \includegraphics[width=\textwidth]{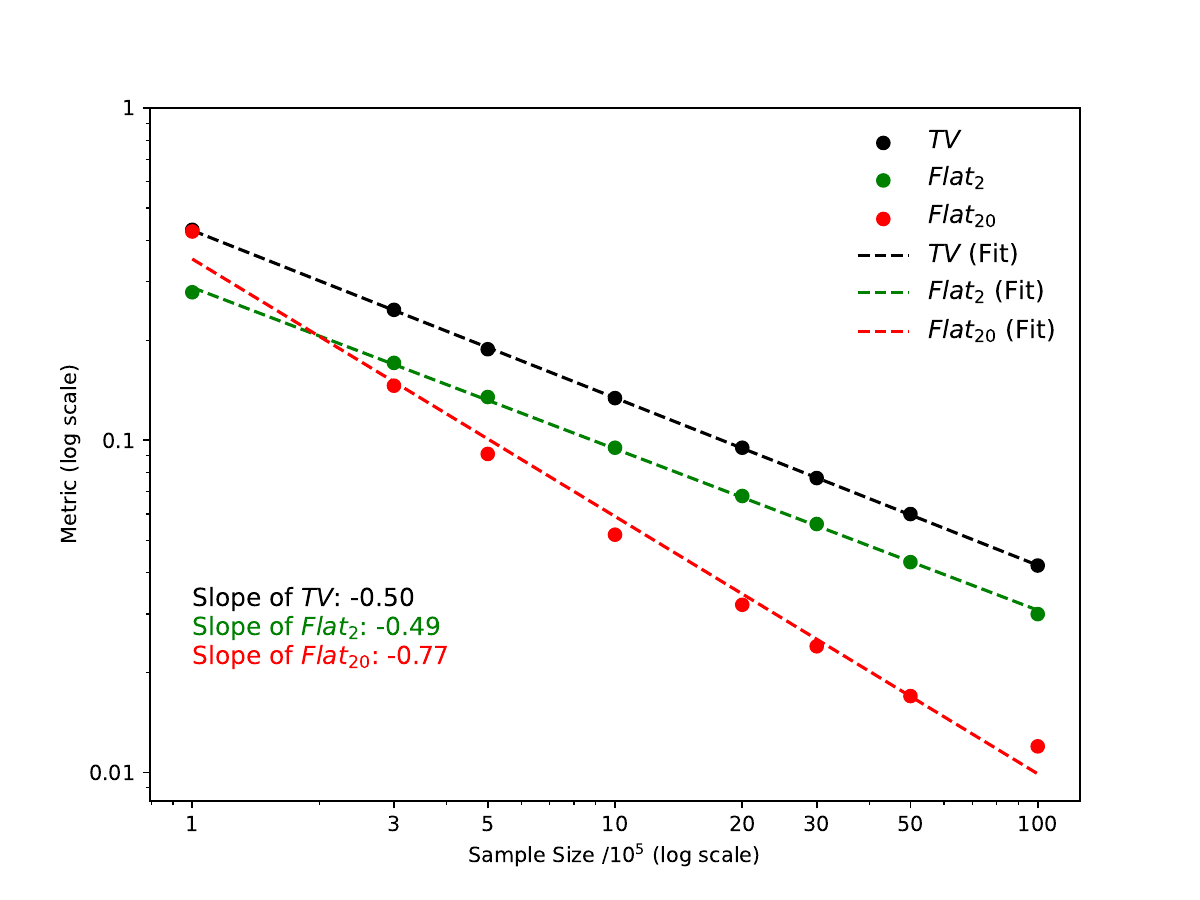}
        \caption*{List}
    \end{minipage}
    \hfill
    \begin{minipage}{0.32\textwidth}
        \centering
        \includegraphics[width=\textwidth]{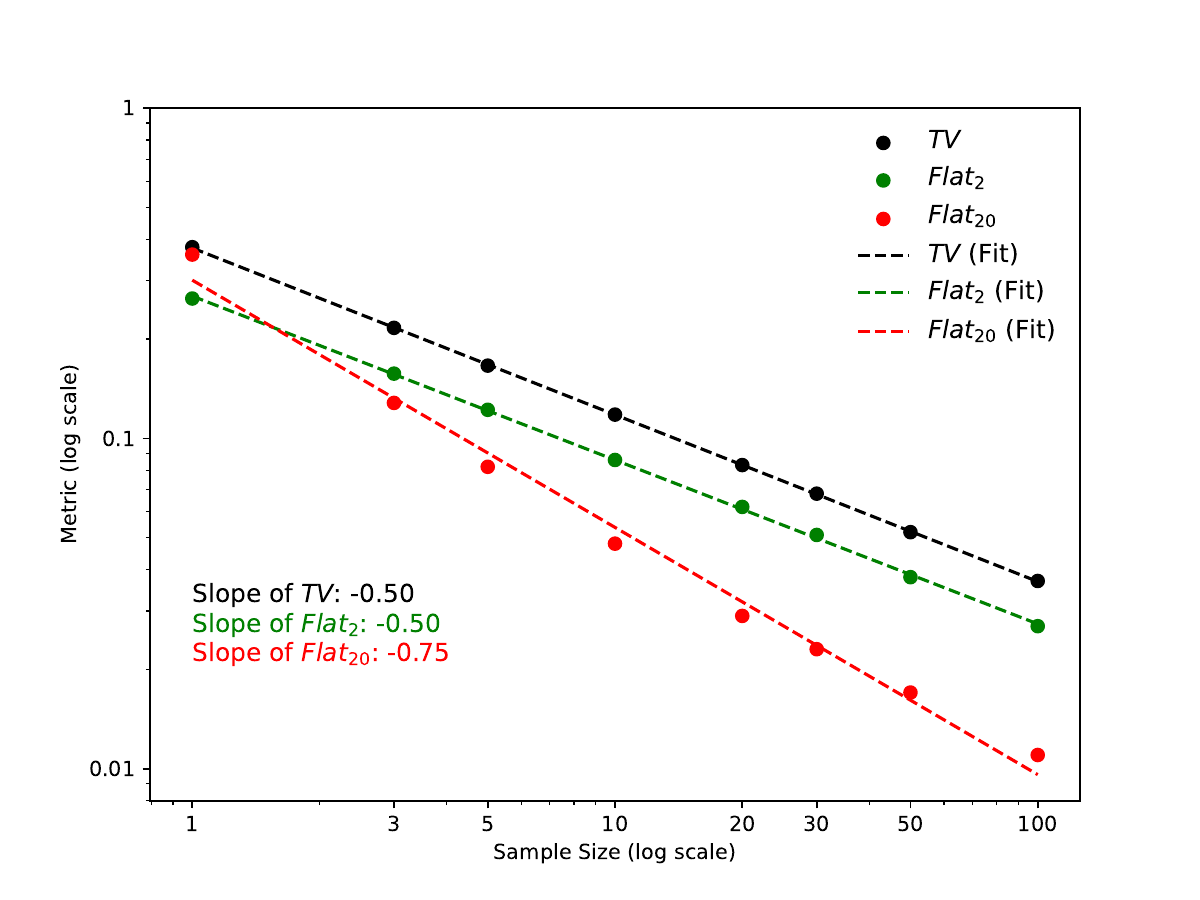}
        \caption*{PCFG}
    \end{minipage}
    \hfill
    \begin{minipage}{0.32\textwidth}
        \centering
        \includegraphics[width=\textwidth]{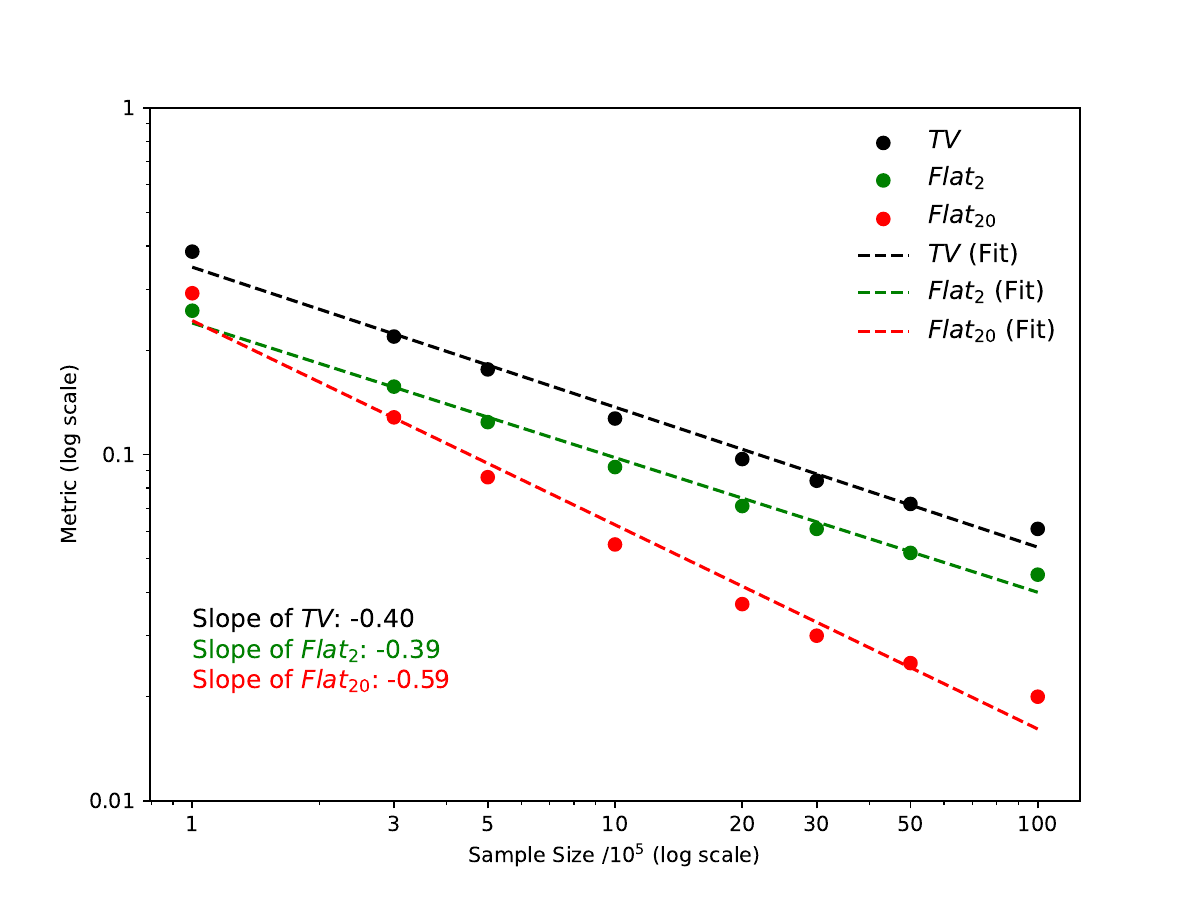}
        \caption*{Markov}
    \end{minipage}
    \caption{Log-Log Plot of Sample Size vs Metric with Linear Regression (12306, $|D|=1.3\times 10^5$)}
    \label{fig_12306}
\end{figure*}

}

\end{document}